\documentclass[a4paper,fleqn,usenatbib]{mnras}

\usepackage[T1]{fontenc}
\usepackage{ae,aecompl}

\usepackage{graphicx}	
\usepackage{amsmath}	
\usepackage{amssymb}	
\usepackage{epsfig}

\def\msol{\hbox{\kern 0.20em $M_\odot$}}
\def\lsol{\hbox{\kern 0.20em $L_\odot$}}
\def\rsol{\hbox{\kern 0.20em $R_\odot$}}
\def\sr{\hbox{\kern 0.20em sr}}
\def\srmu{\hbox{\kern 0.20em sr$^{-1}$}}
\def\g{\hbox{\kern 0.20em g}}
\def\gmu{\hbox{\kern 0.20em g$^{-1}$}}
\def\kg{\hbox{\kern 0.20em kg}}
\def\pc{\hbox{\kern 0.20em pc}}
\def\mum{\hbox{\kern 0.20em $\mu$m}}
\def\mumd{\hbox{\kern 0.20em $\mu$m$^{-2}$}}
\def\cm{\hbox{\kern 0.20em cm}}
\def\m{\hbox{\kern 0.20em m}}
\def\km{\hbox{\kern 0.20em km}}
\def\nm{\hbox{\kern 0.20em nm}}
\def\s{\hbox{\kern 0.20em s}}
\def\h{\hbox{\kern 0.20em h}}
\def\sec{\hbox{\kern 0.20em sec}}
\def\min{\hbox {\kern 0.20em min}}
\def\smu{\hbox{\kern 0.20em s$^{-1}$}}
\def\smd{\hbox{\kern 0.20em s$^{-2}$}}
\def\an{\hbox{\kern 0.20em an}}
\def\anmu{\hbox{\kern 0.20em an$^{-1}$}}
\def\deg{\hbox{\kern 0.20em $^{\rm o}$}}
\def\yr{\hbox{\kern 0.20em yr}}
\def\yrmu{\hbox{\kern 0.20em yr$^{-1}$}}
\def\Myr{\hbox{\kern 0.20em Myr}}
\def\Mymu{\hbox{\kern 0.20em Myr$^{-1}$}}
\def\K{\hbox{\kern 0.20em K}}
\def\pcmu{\hbox{\kern 0.20em pc$^{-1}$}}
\def\pcmd{\hbox{\kern 0.20em pc$^{-2}$}}
\def\pcmt{\hbox{\kern 0.20em pc$^{-3}$}}
\def\kms{\hbox{\kern 0.20em km\kern 0.20em s$^{-1}$}}
\def\kmpd{\hbox{\kern 0.20em km$^{2}$}}
\def\kpc{\hbox{\kern 0.20em kpc}}
\def\cms{\hbox{\kern 0.20em cm\kern 0.20em s$^{-1}$}}
\def\erg{\hbox{\kern 0.20em erg}}
\def\ergs{\hbox{\kern 0.20em erg}}
\def\cmpd{\hbox{\kern 0.20em cm$^2$}}
\def\cmmd{\hbox{\kern 0.20em cm$^{-2}$}}
\def\cmms{\hbox{\kern 0.20em cm$^{-6}$}}
\def\cmpt{\hbox{\kern 0.20em cm$^3$}}
\def\cmmt{\hbox{\kern 0.20em cm$^{-3}$}}
\def\mpd{\hbox{\kern 0.20em m$^2$}}
\def\mmd{\hbox{\kern 0.20em m$^{-2}$}}
\def\mpt{\hbox{\kern 0.20em m$^3$}}
\def\mmt{\hbox{\kern 0.20em m$^{-3}$}}
\def\mujy{\hbox{\kern 0.20em $\mu$Jy}}
\def\mjy{\hbox{\kern 0.20em mJy}}
\def\Mj{\hbox{\kern 0.20em MJy}}
\def\jy{\hbox{\kern 0.20em Jy}}
\def\ghz{\hbox{\kern 0.20em GHz}}
\def\srmd{\hbox{\kern 0.20em sr$^{-1}$}}

\def \mum{$\mu$m}
\def\G{\hbox{\kern 0.20em G}}

\def\htwo{\hbox{H${}_2$}}
\def\h13cop{\hbox{H$^{13}$CO$^{+}$}}

\def\h2o{\hbox{H$_2$O}}

\title[HCN/HNC chemistry in L1157]{HCN/HNC chemistry in shocks: a study of L1157-B1 with ASAI}

\author[]{
B. Lefloch$^{1}$,
G. Busquet$^{1,2,3}$,
S. Viti$^{4,5}$,
C. Vastel$^6$,
E. Mendoza$^{7}$,
M. Benedettini$^8$,
C. Codella$^{9,1}$,
\newauthor
L. Podio $^9$,
A. Schutzer$^{1}$,
P.R. Rivera-Ortiz$^{1}$,
J.R.D. Lépine$^7$, 
R. Bachiller$^{10}$
\\
$^{1}$Univ. Grenoble Alpes, CNRS, IPAG, F-38000 Grenoble, France\\
$^2$Institut de Ciències de l’Espai (ICE, CSIC), Can Magrans, s/n, 08193 Cerdanyola del Vallès, Catalonia, Spain\\
$^3$Institut d’Estudis Espacials de Catalunya (IEEC), 08034 Barcelona, Catalonia, Spain\\
$^{4}$ Leiden Observatory, Leiden University, PO Box 9513, NL-2300 RA Leiden, the Netherlands \\
$^{5}$Department of Physics and Astronomy, University College London, Gower Street, London, WC1E 6BT, England \\
$^6$IRAP, Université de Toulouse, CNRS, UPS, CNES, 31400 Toulouse, France\\
$^{7}$Instituto de Astronomia, Geofísica e Ciências Atmosféricas, Universidade de São Paulo, São Paulo 05508-090, SP, Brazil\\
$^{8}$INAF, Istituto di Astrofisica e Planetologia Spaziali, via Fosso del Cavaliere 100, 00133 Roma, Italy\\
$^{9}$INAF, Osservatorio Astrofisico di Arcetri, Largo Enrico Fermi 5, I-50125 Firenze, Italy\\ 
$^{10}$IGN, Observatorio Astron\'omico Nacional, Calle Alfonso XII, 3 E-28004 Madrid, Spain\\
}

\date{Accepted 2021 July 13. Received 2021 June 14; in original form 2021 April 14.}

\pubyear{2021}

\begin{document}
\label{firstpage}
\pagerange{\pageref{firstpage}--\pageref{lastpage}}
\maketitle

%
\begin{abstract}
HCN and its isomer HNC play an important role in molecular cloud chemistry and the formation of more complex molecules. We investigate here the impact of protostellar shocks on the HCN and HNC abundances from high-sensitivity IRAM 30m observations of  the prototypical shock region L1157-B1  and the envelope of the associated Class 0 protostar, as a proxy for the pre-shock gas. The isotopologues H$^{12}$CN, HN$^{12}$C, H$^{13}$CN, HN$^{13}$C, HC$^{15}$N, H$^{15}$NC, DCN and DNC were all detected towards both regions. Abundances and excitation conditions were obtained from  radiative transfer analysis of molecular line  emission under the assumption of Local Thermodynamical Equilibrium. In the pre-shock gas, the abundances of the HCN and HNC isotopologues are similar to those encountered in dark clouds, with a HCN/HNC  abundance ratio $\approx 1$ for all isotopologues. A strong D-enrichment (D/H$\approx 0.06$) is measured in the pre-shock gas. There is no evidence  of $^{15}$N fractionation neither in the quiescent nor in the shocked gas.  At the passage of the shock, the HCN and HNC abundances increase in the  gas phase in different manners so that the HCN/HNC relative abundance ratio increases by a factor 20.  The gas-grain chemical and shock model UCLCHEM allows us to reproduce the observed trends for a C-type shock with pre-shock density $n$(H)= $10^5\cmmt$  and shock velocity $V_s= 40\kms$. We conclude that the HCN/HNC variations across the shock are mainly caused by the  sputtering of the grain mantle material in relation with the history of the grain ices. 
\end{abstract}



\section{Introduction}
\label{introduction}
Hydrogen cyanide (HCN) is one of the most simple interstellar molecules. Thanks to its large dipole moment (2.99 Debye; \citealt{Bhattacharya1960}), its rotation transitions are good probe of dense molecular gas in Galactic and extragalactic environments. 
HCN and its isomer hydrogen isocyanide HNC are thought to play an important role in the formation of more complex molecules, like cyanopolyynes HC$_{2n+1}$N, either in dark cloud cores \citep{Suzuki1992} or in more energetic regions, like protostellar shocks \citep{Mendoza2018}. Since HCN and HNC have similar energy spectra and dipole moments, their differences in spatial distribution is mainly related to the gas chemical conditions.  For this reason, they have often been used as probes of gas chemical evolution in both dense cores and 
star-forming regions \citep{Schilke1992,Ungerechts1997,Daniel2013}. 

Many systematic studies of dark clouds and low-mass protostellar cores  have shown  the HNC/HCN ratio to be close to unity \citep[see e.g.,][]{Irvine1984,Hirota1998,hily2010}. No difference is observed between the values measured in prestellar and protostellar cores, implying that the evaporation of HCN and HNC from dust grains does not contribute significantly to the observed emission in the cold envelope. On the contrary, towards the high-mass star forming region OMC-1 in Orion, the HNC/HCN ratio  displays strong variations with especially low values $\approx 0.01$ towards the hot core regions  while it is of the order of  0.2 in adjacent ridge positions. While the abundance of HCN is similar to that of dark cloud cores, the HNC abundance is 2 orders of magnitude lower in the high-temperature gas of the hot core \citep{Schilke1992}. A somewhat similar behaviour is observed  towards IRAS16293$-$2422 when looking at the high-excitation lines of HCN and HNC \citep{vanDishoeck1995}.  
Recently, \citet{Hacar2020} demonstrated the high sensitivity of the HCN/HNC  $J$=1--0 line intensity ratio to the gas kinetic temperature.  

In his pioneering work on outflow shock chemistry \citet{Bachiller1997} brought the first hints of HCN abundance enhancement  in  protostellar shock region. At that time, only a low number of molecular transitions was observed with the IRAM 30m telescope, hence preventing an accurate determination of the excitation conditions and molecular abundances. From a theoretical point of view, both the dependence of the HCN/HNC ratio to the temperature and the sputtering of dust grains are two processes which could {\em a priori} alter molecular gas abundances of HCN and HNC across a shock. 

Many subsequent observational studies on the physical and chemical characterization of L1157-B1 have been carried out both with (sub)millimeter single-dish and  interferometers, which led to a rather detailed, consistent picture of the outflow shock while unveiling its chemical richness (see e.g. \citet{Codella2010,Codella2017,Viti2011,Lefloch2012,Lefloch2016,Lefloch2017,Busquet2014,Mendoza2014,Mendoza2018,Podio2014, Podio2016,Podio2017,Gomez2015}). 
The spatial distribution of the HCN emission in the L1157 southern outflow lobe was obtained at $\approx 5\arcsec$ scale  for the first time by \citet{Benedettini2007} with the IRAM Plateau de Bure interferometer.  The emission of the  rare  $^{13}$C, D and $^{15}$N isotopologues was investigated  a few years later at a few arcsec resolution  with the IRAM NOEMA interferometer \citep{Busquet2017,Benedettini2021}. In particular, the distribution of DCN  was shown to result from  a combination of gas-phase chemistry that produces the widespread DCN emission, dominating especially in the head of the bow-shock, and sputtering from grain mantles toward the jet impact region. 

These results incited us to revisit  the emission of  the HCN and HNC isotopologues  in  L1157-B1 in a more comprehensive and accurate way than was possible before. This study benefits from  the unbiased  and high-sensitivity millimeter spectral line survey of the shock region carried out as part of the IRAM 30m Large Program ASAI ("Astrochemical Surveys At IRAM", \citet{Lefloch2018}). In order to better understand and constrain more precisely the impact of the shock on the chemistry of  HCN and HNC, we have also investigated the properties of their isotopologue emission in the envelope of the Class 0 protostar L1157-mm, at about 1~arcmin away,  which we took as a proxy of the initial gas composition before the arrival of the shock.  The origin of the molecular emission was then derived  from comparison of our observational results with the  predictions of the time-dependent gas-grain chemical and shock model UCLCHEM \citep{Holdship2017}.  

The paper has been organized as follows. In Sect.~2, we summarize the main observational properties of the shock region L1157-B1. The observations are described in Sect.~3. In Sect.~4, we present our results on the HCN spatial distribution in the region and the gas properties (excitation temperature, column density, abundance) obtained for the HCN and HNC isotopologues based on a simple radiative transfer analysis of the line emission. We discuss in Section~5 the behaviour of HCN and HNC across the shock and we analyse the origin of the molecular emission from comparison with the results of our modelling.  Finally, we present our conclusions in Sect.~6.

\section{The source}
\label{source}

The GAIA mission has led to a revision of the distance to L1157 by several groups \citet{Dzib2018} and \citet{Sharma2020}, who found $(360\pm 32)\pc$ and $(340\pm 21)\pc$, respectively. In this work, we will adopt the value of  $(352\pm 19)\pc$ derived by \citet{Zucker2019}, in agreement with \cite{Benedettini2021}. 
\citet{Gueth1996,Gueth1998} have  studied at high-angular resolution ($\approx 3\arcsec$)  the  structure and dynamics
of the southern lobe of the outflow driven by the protostar L1157-mm from  the emission of the CO $J$=1--0 and SiO $J$=2--1 lines observed with the IRAM Plateau de Bure interferometer. 
These authors showed that the southern lobe of this molecular outflow consists of two cavities, likely created by the propagation of large bow shocks due to episodic events in a precessing, highly collimated, high-velocity jet. This jet was detected by \citet{Tafalla2015} and imaged by \citet{Podio2016}. Based on higher-angular and high-sensitivity CO observations with NOEMA at $0.3\arcsec$, \citet{Podio2016} refined the jet precession modelling by \citet{Gueth1996} and estimated an age of $\approx 1500\yr$ for B1 (and $2500\yr$ for B2), adopting the new distance of $352\pc$. 
Located at the apex of the more recent cavity, the bright bow shock region B1 displays a peculiar molecular complexity \citep[e.g.,][]{Codella2010,Benedettini2012}, which makes it a benchmark for magnetized shock models \citep[e.g.,][]{Gusdorf2008a,Gusdorf2008b,Viti2011}.
Multi-transition analysis of the emission of tracers such as CO \citep{Lefloch2010,Lefloch2012}, H$_2$O \citep{Busquet2014} and CS \citep{Gomez2015}
has allowed to elucidate a coherent scenario where the molecular emission appears to arise from
four physically distinct components, with specific excitation conditions:
\begin{itemize}
\item~Component $g_1$: \citet{Lefloch2012} and \citet{Benedettini2012} evidenced a region of high excitation of $\approx 10\arcsec$ size, with a kinetic temperature 
$T_{\rm kin}\approx 200\K$ and gas density $n(\htwo)\simeq 10^6\cmmt$. This region is associated with the impact of the jet against the L1157-B1 bow shock.
\item~Component $g_2$ is tracing  the outflow cavity associated with L1157-B1, for which a kinetic temperature  $T_{\rm kin} \approx  60\K$ and a gas density $n(\htwo)\simeq$ (1--10)$\times 10^5\cmmt$ were estimated. 
\item~Component $g_3$  is tracing the late outflow cavity associated with L1157-B2, for which have been estimated $T_{\rm kin}\approx 20\K$ and $n(\htwo)\simeq 10^5\cmmt$
\item~ A hot ($T_{\rm kin}\simeq 1000\K$) and tenuous gas component ($n(\htwo)\simeq 10^3-10^4\cmmt$) with a size of $2\arcsec$ - 
\end{itemize}


\section{Observations}
\label{observations}

\subsection{ASAI}

Observations of L1157-B1 and L1157-mm were carried out during several runs between September 2012 and  March 2015 as part of the Large Program ASAI \citep{Lefloch2018}.  The source nominal positions  are $\alpha_{J 2000} =$ 20$^{\text h}$ 39$^{\text m}$ 10.$^{\text s}$2 
$\delta_{J 2000} =$ +68$^{\circ}$ 01$^{\prime}$ 10$^{\prime\prime}$ for L1157-B1 and  $\alpha_{J 2000} =$ 20$^{\text h}$ 39$^{\text m}$ 06.$^{\text s}$3 $\delta_{J 2000} =$ +68$^{\circ}$ 02$^{\prime}$ 15.8$^{\prime\prime}$ for L1157-mm. 
We made use of the broad-band EMIR receivers connected to Fast Fourier Transform Spectrometers in the 200 kHz spectral resolution mode. 

Detailed information on the observation and the data reduction procedures can be found in \citet{Lefloch2018}. 
The ASAI observations were carried out using the "Wobbler Switching" (WSW) mode with a throw of $3\arcmin$.


\subsection{On-The-Fly mapping}
The ASAI dataset was complemented with  a fully sampled map of the HCN $J$= 3--2 transition  in the southern lobe of the L1157 outflow with the IRAM~30m telescope in December 2013. The size of the map, $2\arcmin \times 2.5\arcmin$, is  large enough to encompass the whole southern lobe.  Observations
were carried out with the receiver E230 and the FTS spectrometer in its 192~kHz resolution mode, using the On-The-Fly mode.
We chose a reference position offset by $10\arcmin$ in right ascension from L1157-B1, and checked to be free of emission in
the HCN $J$=3--2 line.  Atmospheric calibrations were
performed every 15~min and showed the weather to be stable, with 2.5 -- 3.0~mm of precipitable water vapour, and system
temperatures in the range 450 -- 550~K.  Pointing was monitored every  hour and corrections were always found lower than $3\arcsec$.

\subsection{Data reduction}

\begin{figure*}
\centering{
\includegraphics[width=2\columnwidth,keepaspectratio]{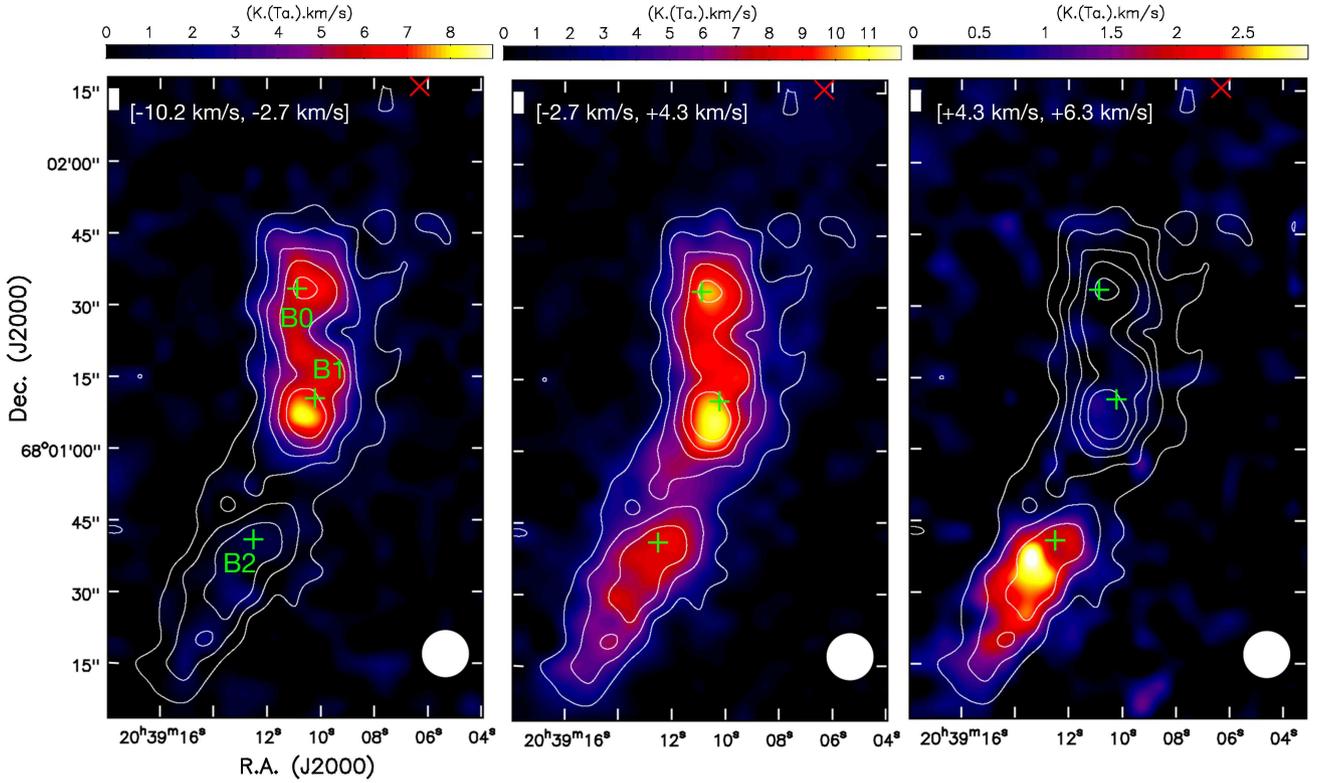}
\caption{HCN $J$=3--2 velocity-integrated emission in the southern lobe of the L1157 outflow. The emission is integrated over the velocity range indicated in each panel (colour scale)  to illustrate the separation of the emitting gas into the B0 and B1 shocks and the B2 shock. The emission in the central panel between $-2.7$ and $+4.3\kms$ encompasses the emission from the ambient gas  at  $V_{lsr}$= $+2.7\kms$. 
White contours represent the total velocity-integrated emission of the HCN $J$=3--2 line. Contour levels range from 3$\sigma$ to 15$\sigma$ in steps of of 3$\sigma$, where $\sigma=1.07$~K\,km\,s$^{-1}$ is the rms noise of the map. In each panel, green crosses mark  the nominal position of shocks B0, B1, and B2; a red cross marks the nominal position of protostar L1157-mm  \citep{Gueth1997}. The size (HPBW) of the telescope main beam is draw by the white circle in the bottom right corner of each panel. }
\label{map-hcn}
}
\end{figure*}

\begin{figure}
\centering{
\includegraphics[width=0.9\columnwidth]{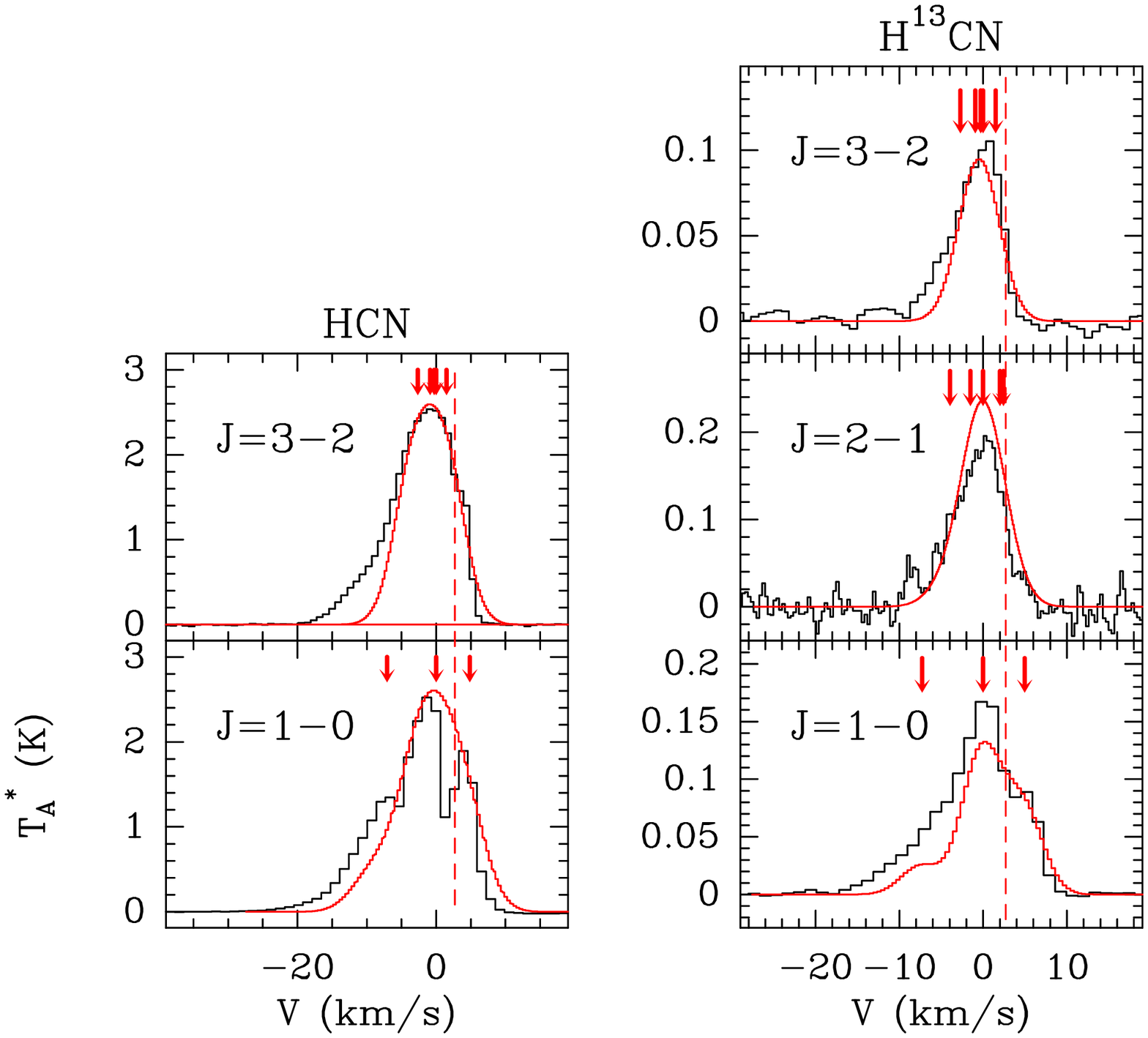} 
\includegraphics[width=0.9\columnwidth]{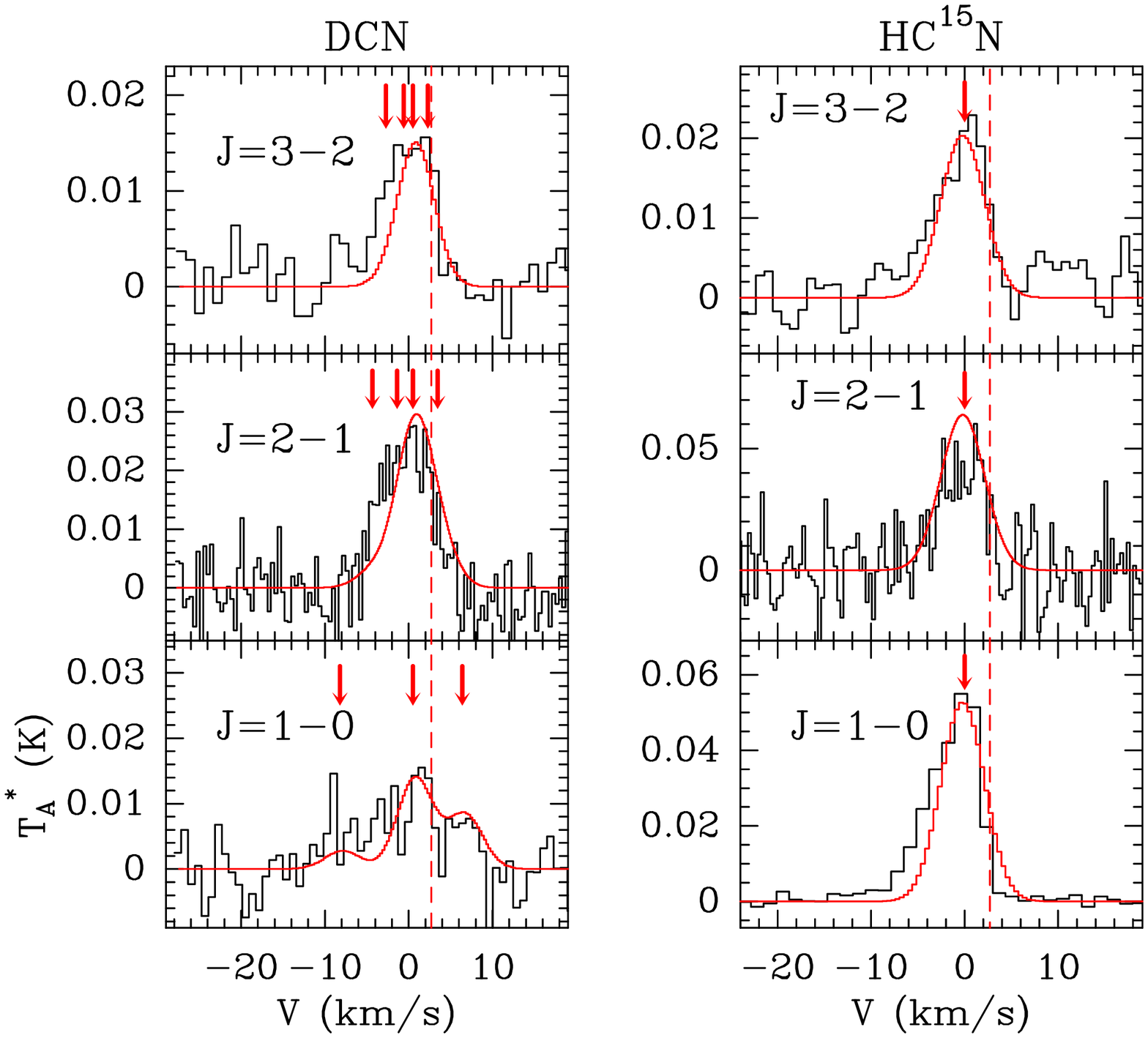}
\caption{Montage of line profiles  of  HCN and its rare isotopologues (H$^{13}$CN, DCN, HC$^{15}$N), as  observed with ASAI towards L1157-B1. Dashed-dotted lines indicate the baseline and rest velocity, $V_{lsr}$= $+2.7\kms$. All the spectra are smoothed to a resolution of $1\kms$. The best fits obtained from the  LTE analysis with CASSIS are  displayed in red. The velocity axis is associated to the reference frequency given in Table~2. Red arrows mark the location of the hyperfine satellites on the velocity axis, based on the frequencies given in the CASSIS database. 
\label{l1157b1-hcn}}}
\end{figure}

\begin{figure}
\centering{
\includegraphics[width=0.9\columnwidth]{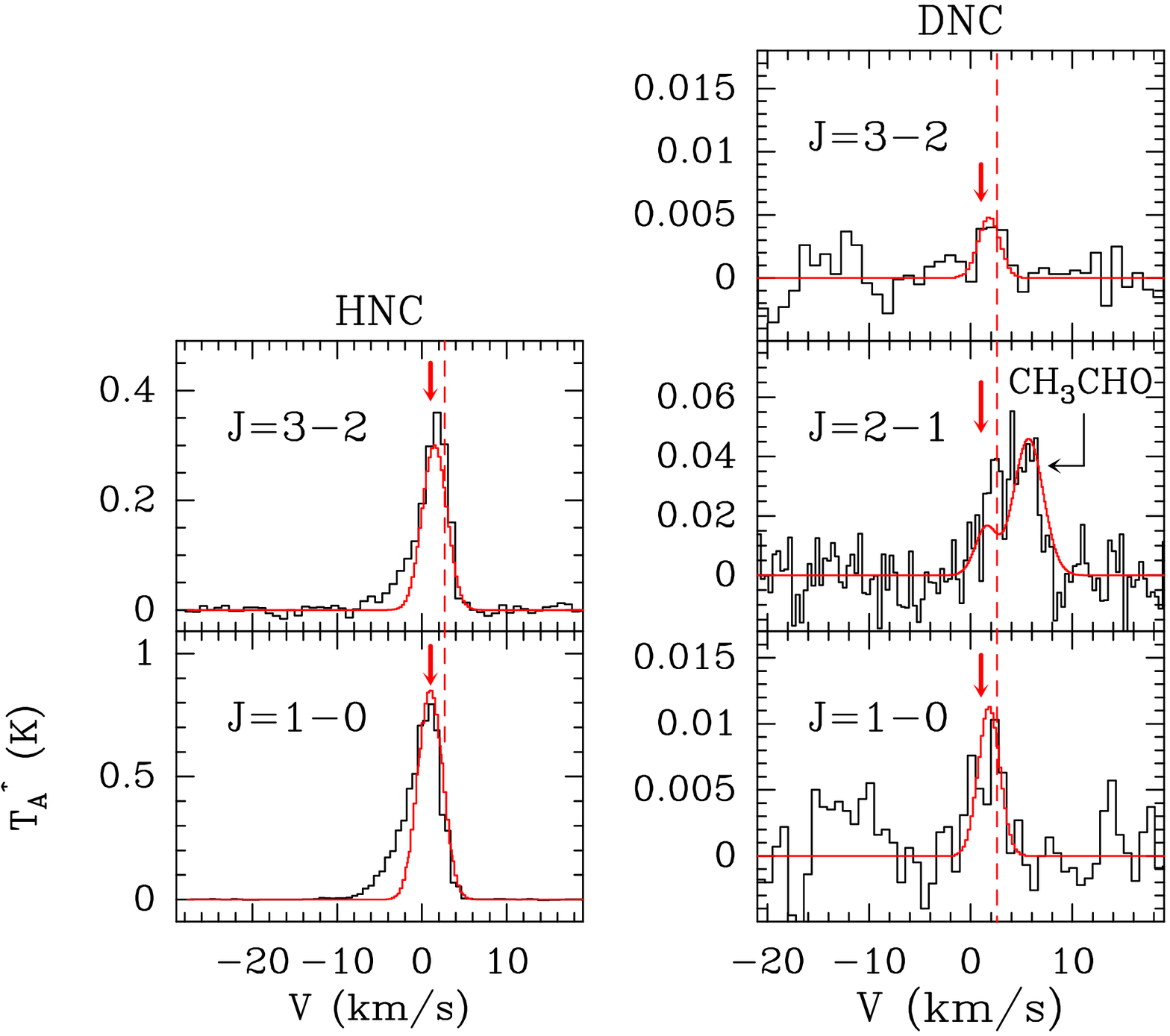} 
\includegraphics[width=0.9\columnwidth]{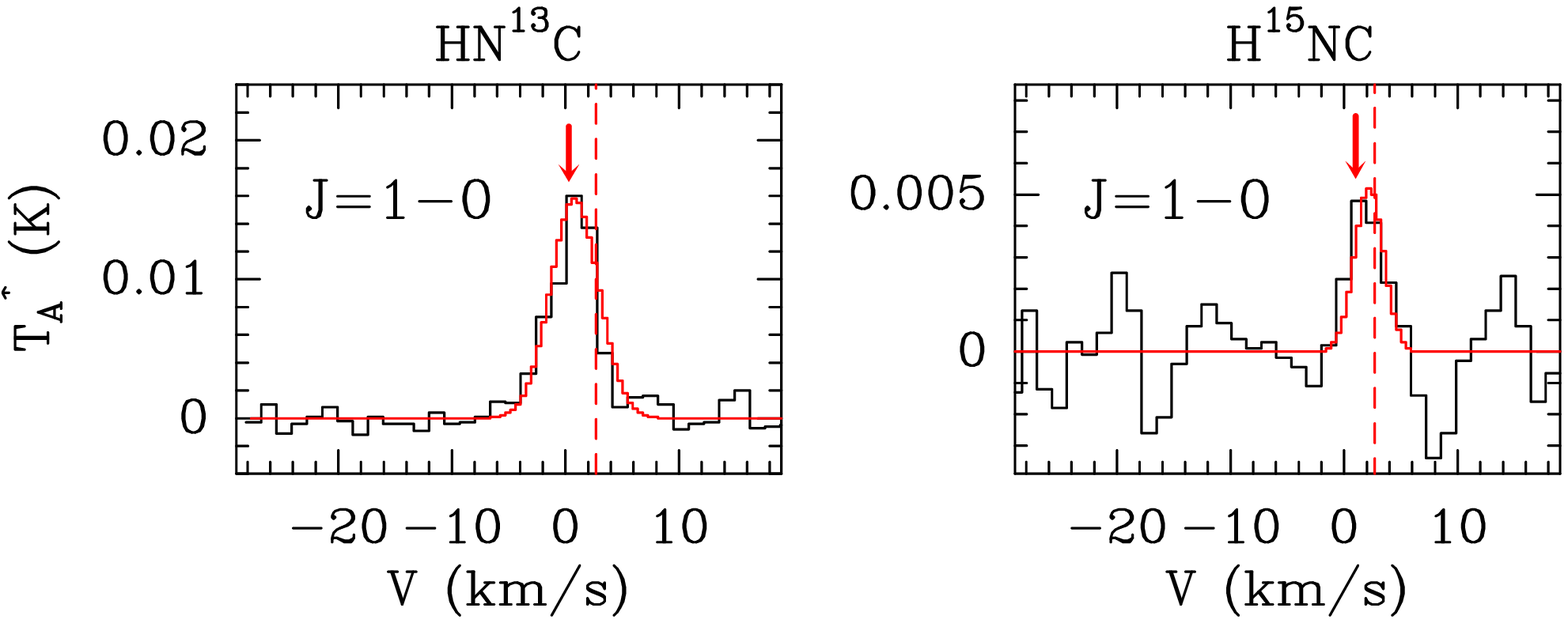}
\caption{Montage of line profiles  of  HNC and its rare isotopologues (HN$^{13}$C, DNC, H$^{15}$NC), as  observed with ASAI  towards L1157-B1. Dashed-dotted lines indicate the baseline and rest velocity, $V_{lsr}$= $+2.7\kms$. All the spectra are smoothed to a resolution of $1\kms$. The best fits obtained from the LTE analysis with CASSIS are  displayed in red. The velocity axis is associated to the reference frequency given in Table~2. Red arrows mark the location of the hyperfine satellites on the velocity axis, based on the frequencies given in the CASSIS database. 
\label{l1157b1-hnc}}}
\end{figure}

\begin{figure}
\centering{
\includegraphics[width=0.9\columnwidth]{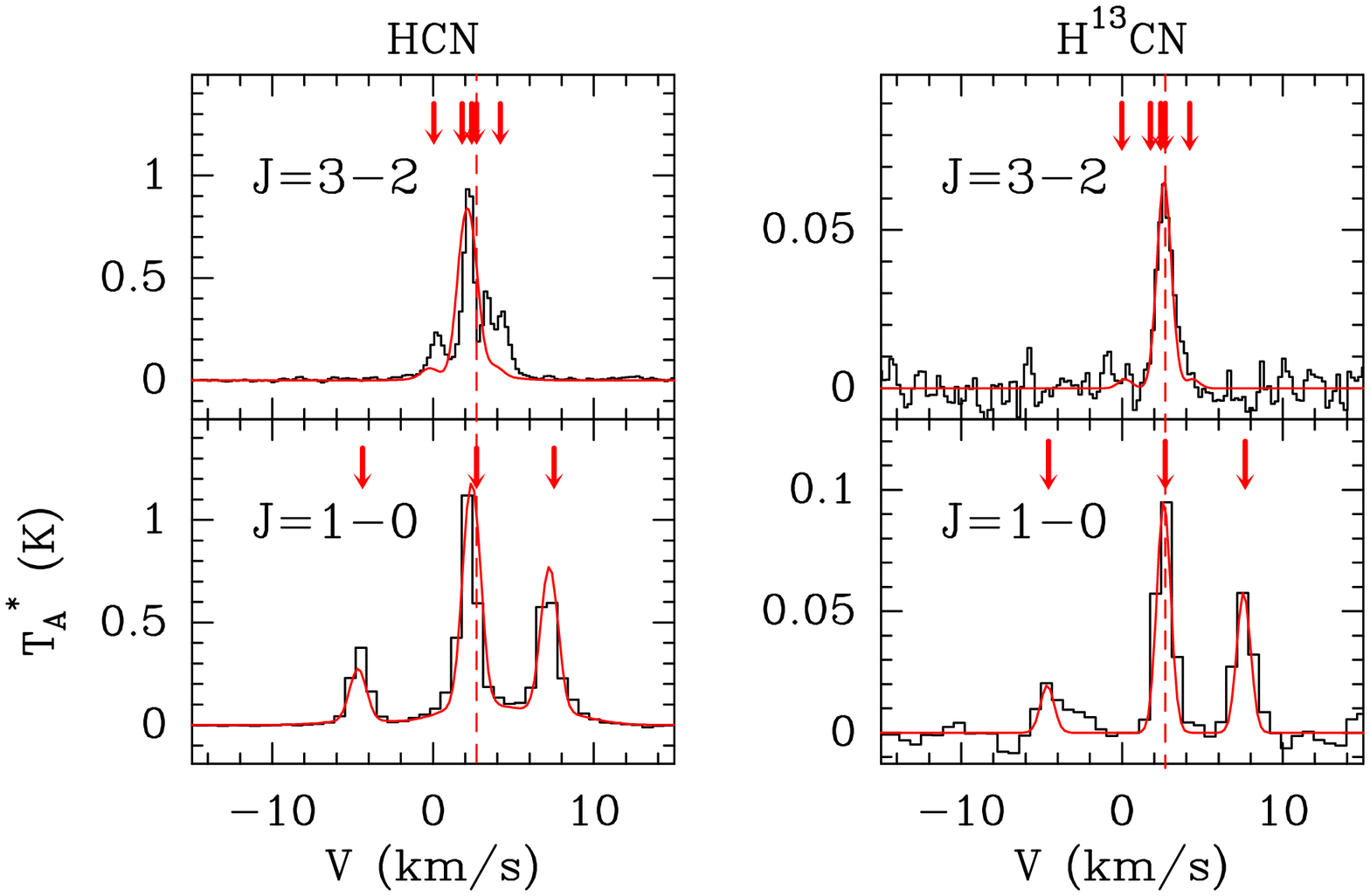} 
\includegraphics[width=0.9\columnwidth]{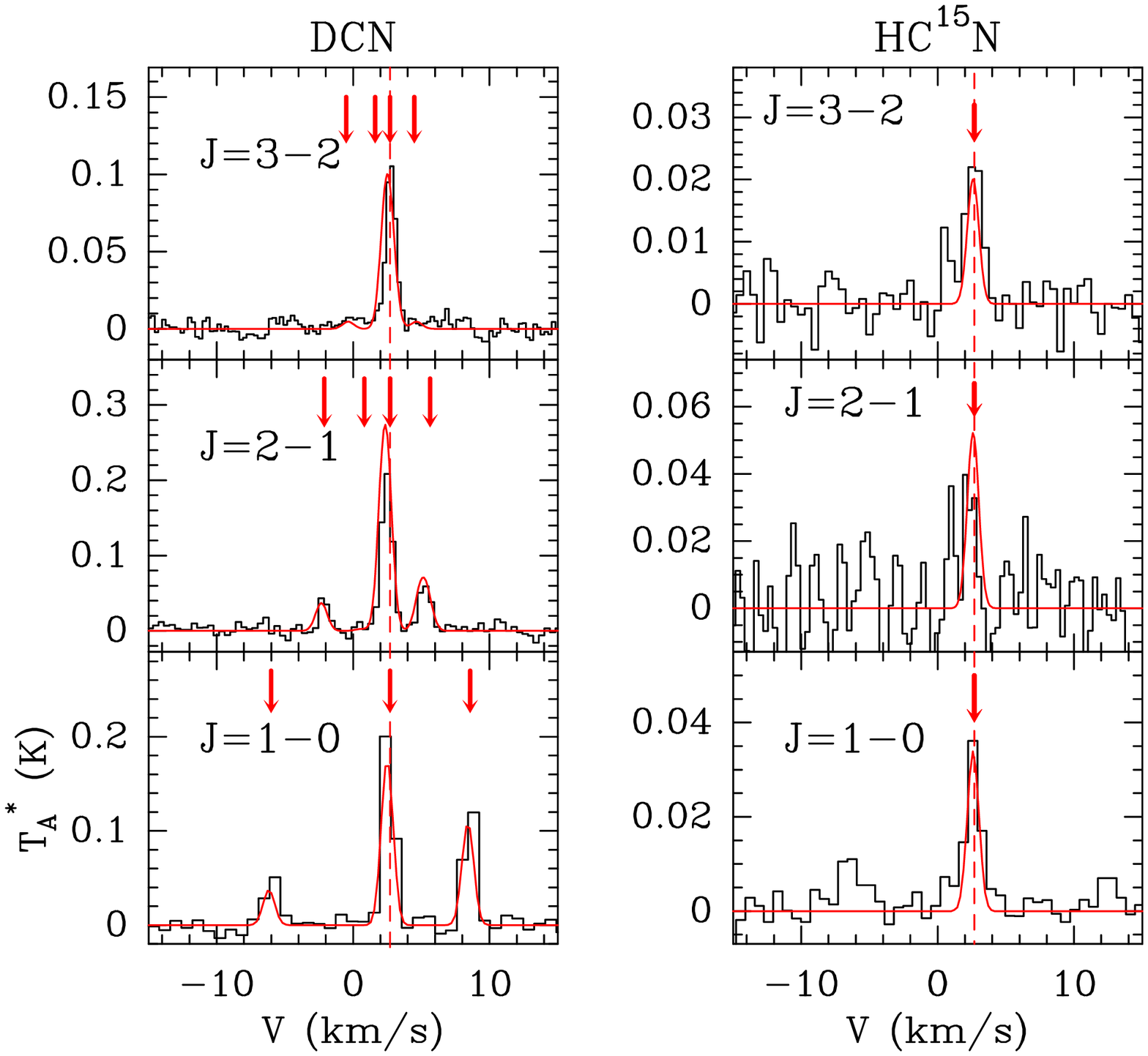}
\caption{Montage of line profiles  of  HCN and its rare isotopologues (H$^{13}$CN, DCN, HC$^{15}$N), as  observed with ASAI towards L1157-mm. Dashed-dotted lines indicate the baseline and rest velocity, $V_{lsr}$= $+2.7\kms$. The best fits obtained from the  LTE analysis with CASSIS are  displayed in red. The velocity axis is associated to the reference frequency given in Table~2. Red arrows mark the location of the hyperfine satellites on the velocity axis, based on the frequencies given in the CASSIS database. 
\label{l1157mm-hcn}}}
\end{figure}

\begin{figure}
\centering{
\includegraphics[width=0.9\columnwidth]{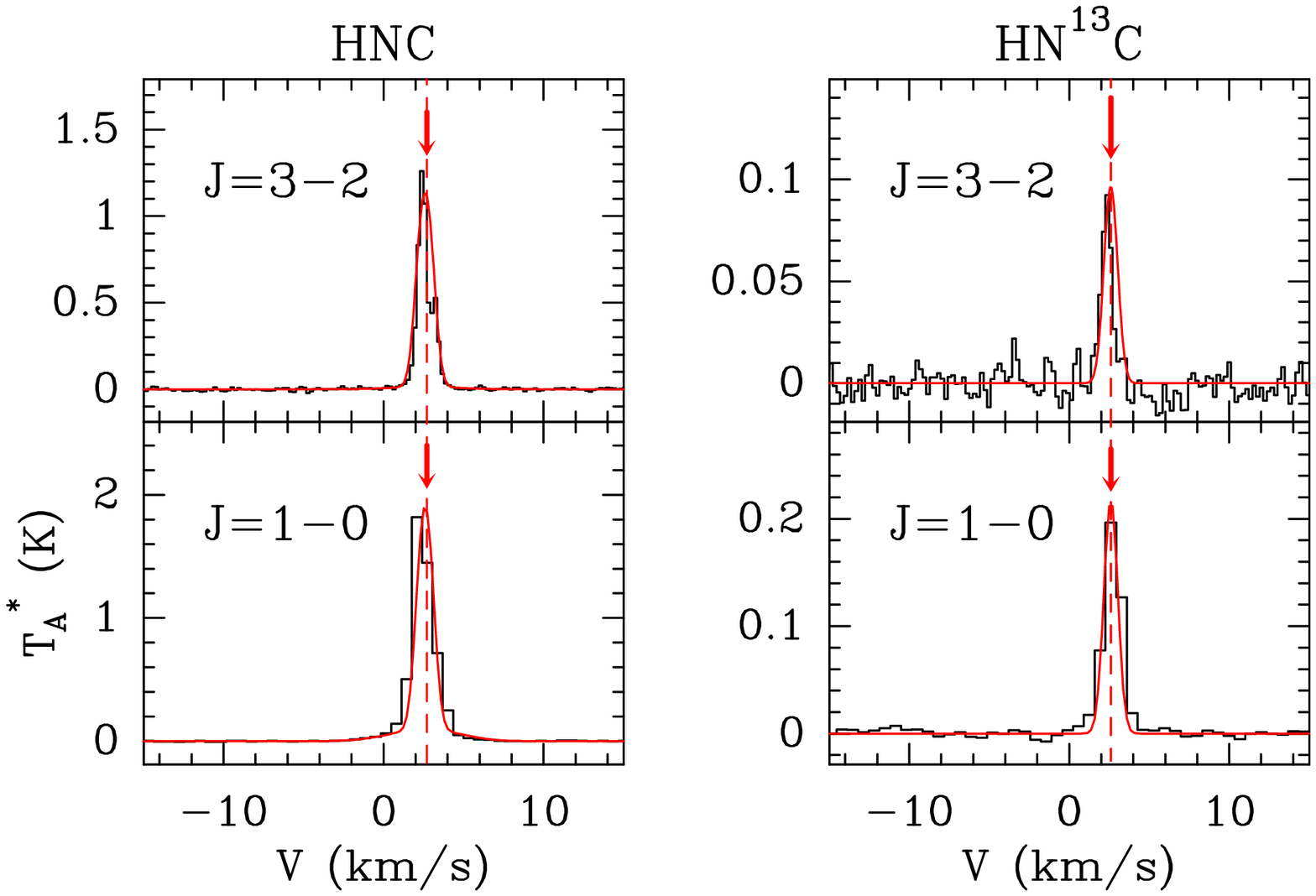} 
\includegraphics[width=0.9\columnwidth]{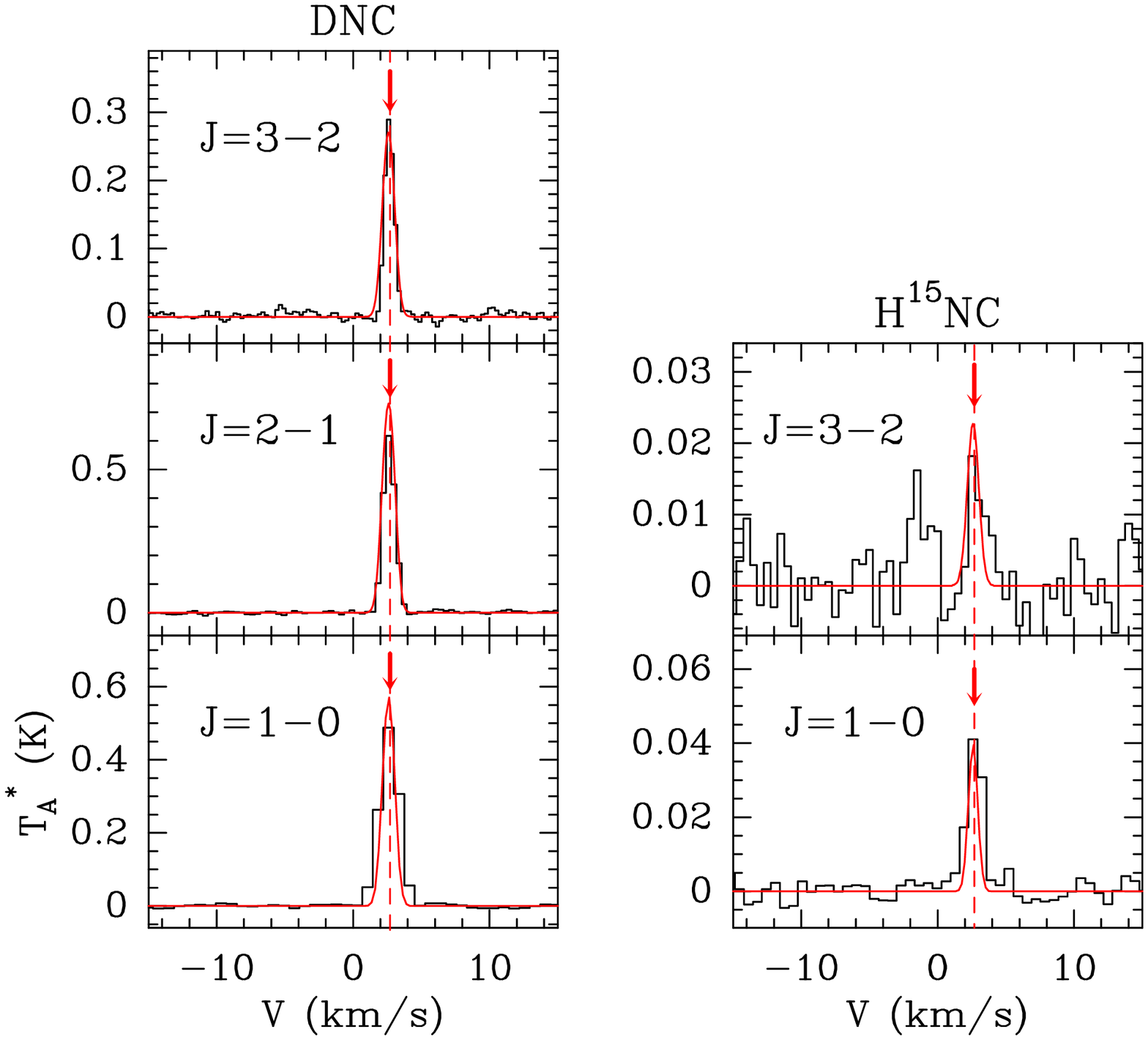}
\caption{Montage of line profiles  of  HNC and its rare isotopologues (HN$^{13}$C, DNC, H$^{15}$NC), as  observed with ASAI towards L1157-mm. Dashed-dotted lines indicate the baseline and rest velocity, $V_{lsr} = +$2.7~km s$^{-1}$. The best fit obtained from the  LTE analysis with CASSIS are  displayed in red. The velocity axis is associated to the reference frequency given in Table~2. Red arrows mark the location of the hyperfine satellites on the velocity axis, based on the frequencies given in the CASSIS database. 
\label{l1157mm-hnc}}}
\end{figure}

The data reduction was performed using the GILDAS/CLASS software\footnote{https://www.iram.fr/IRAMFR/GILDAS/}. 
We have detected several transitions of HCN, HNC, and their rare $^{13}$C-, D- , $^{15}$N-  isotopologues, between 72 and 272 GHz,  towards the shock region L1157-B1 and the protostar L1157-mm. The line spectra are presented in Figs.~\ref{l1157b1-hcn}--\ref{l1157mm-hnc}.  
The resolution of the spectrometer  allowed us to resolve the hyperfine structure of the rotational transitions of HCN, H$^{13}$CN and DCN. The location of the hyperfine components are indicated by red arrows  in the spectra.  The line intensities  are expressed in units of antenna temperature corrected for atmospheric attenuation and rearward losses ($T_A^{\ast}$). 

For subsequent analysis and radiative transfer modelling,  fluxes were  expressed in main beam temperature units ($T_{\rm mb}$). The telescope and receiver parameters (beam efficiency, $B_{\mathrm{eff}}$; forward efficiency,
$F_{\mathrm{eff}}$; Half Power beam Width, HPBW) were taken from the IRAM webpage\footnote{http://publicwiki.iram.es/Iram30mEfficiencies/}. The calibration uncertainties are typically 10, 15, 20\% at 3~mm, 2~mm, 1.3~mm, respectively. 

The spectroscopic properties and the observational parameters of all the detected transitions  are summarized in Tables~\ref{tbl-1}--\ref{tbl-2}, for L1157-B1 and L1157-mm, respectively.
We extracted the observational properties of the line profiles (velocity-integrated flux, Full Width at Half Maximum, emission peak velocity) performing a simple gaussian fitting using CLASS whenever possible. The only exceptions concern HCN and H$^{13}$CN in L1157-B1, where the presence of several components prevents from disentangling the emission of the hyperfine satellites.
In that case, the total flux (in $T_A^{\ast}\kms$) was obtained from integrating over the whole velocity range of emission. 

\begin{table*}
\centering{
\caption{Spectroscopic and observational parameters of the transitions of the HCN and  HNC and their rare D, $^{13}$C and $^{15}$N isotopologues detected towards the protostellar shock~L1157-B1.   Note that the flux uncertainties reported here contain only the statistical rms noise. TAG is a 5-digit code which indicates the spectroscopic database used for CASSIS line identification and modelling. }
\label{tbl-1}
\begin{tabular}{llrrrrcclll}
\hline
Species & TAG & Transition & Frequency & $E_u$ & $A_{ul}$ & HPBW & $\eta_{mb}$   & $\int T_{A}^*dv$  & FWHM & $V_{lsr}$ \\
            &         &                 &  MHz &  K  & $10^{-5}$ s$^{-1}$  & $^{\prime\prime}$ &  &   K km s$^{-1}$  & km s$^{-1}$ & km s$^{-1}$ \\
\hline
HCN              &  27501 & 1 -- 0	&	88631.602	&	4.3	&	2.4	&	27.8	&	0.85	  & 31.2(1.5)  &4 .7(1)	    &	-0.2(1.0)	\\
                     &             & 3 -- 2	& 265886.434	&	25.5	&	83.6	&	9.3   &	0.59	  & 29.9(1.0)	   & 9.5(2)	    &	-1.3(.7)	\\
DCN              & 28509 & 1 -- 0   &   72414.905 & 3.5   & 1.33  &  34.0 & 0.86   & 0.12(.03)  & 7.9(.5)   & -0.2 (.1) \\
                     &            & 2 -- 1   & 144828.001 & 10.4 & 12.7 &  17.0 &  0.79  &  0.19(.01) & 7.0(.3)   & -0.3(.2)   \\
                     &            & 3 -- 2   & 217238.538 & 20.9 & 46.0 &  11.3 &  0.65  & 0.10(.01)   & 6.5(.6)   &-0.1(.3)   \\
H$^{13}$CN  & 28501 & 1 -- 0   & 86340.184	&	4.1	&	2.23	&	28.5	 &	0.85	   &	1.76(.03)	&	4.8(1)	&	-0.4(.2)	\\
                     &            & 2 -- 1	& 172677.851	&	12.4	&	21.4	&	14.3	&	0.75	  &	1.20(.01)	&	7.2(4)	&	-0.3	(1)	\\
                     &            & 3 -- 2	& 259011.798	&	24.9	&	77.3	&	9.5   & 0.60	  &	0.73(.02)	&	6.2(1)	&	-0.34(.3)	\\
HC$^{15}$N  & 28506 & 1 -- 0   & 86054.961   &   4.1 & 2.2    &  28.6 & 0.86   &  0.35(.01)  & 6.1(.1)   & -1.0(.1) \\
                     &            & 2 -- 1   & 172107.956 &  12.4& 21.1  &  14.3 & 0.75   &  0.27(.01)  & 5.6(.6)   & -0.3(.3)  \\
                     &            & 3 -- 2   & 258157.100 &  24.8& 76.5  &  9.5   &  0.60  &  0.13(.01)  & 6.2(.7)  & -0.1(.3)   \\ \hline
HNC              & 27502 & 1 -- 0   & 90663.568	&	4.4 	&2.69	&	27.1  &	0.85   &	4.26(.02) &	4.5 (1) &	0.3(.1)  	\\
                     &            & 3 -- 2	& 271981.142	&	26.1 &	93.4 &	9.0 	 &	0.58   &	2.36(.04) &	3.7(1)  & 1.5(.3)  	\\
DNC              &  28508 &1 -- 0   & 76305.701   &  3.7  & 1.60 & 32.2  & 0.86    & 0.010(.004)&1.4(.6)   & 0.3(.3)     \\
                     &             &2 -- 1   &152609.746  & 11.0 & 15.4 & 16.1  &  0.77  & 0.04(.02)   &1.7(.6)   & 0.6(.4)  \\
                     &             &3 -- 2   & 228910.481& 22.0& 55.7 & 10.7   & 0.64   &  0.012(.003)&2.7(.7)  & 2.1(.3)   \\
HN$^{13}$C  & 28515 & 1 -- 0	&   87090.825	&	4.2	&	2.38	&	28.3 &	0.85  &	0.093(.005)	&5.2(.3)	&	0.8(1)	\\
H$^{15}$NC  & 28006 & 1 -- 0	&	 88865.715 & 4.3    & 2.53 & 27.8  & 0.85  & 0.025(.008) & 3.4(.3)   & 1.8(.1) \\ 
\hline 
\end{tabular}
}
\end{table*}

\begin{table*}
\centering{
\caption{Spectroscopic properties and observational parameters of the transitions of the HCN and  HNC and their rare D, $^{13}$C and $^{15}$N isotopologues detected towards the protostar L1157-mm.  Note that the flux uncertainties reported here contain only the statistical rms noise.  TAG is a 5-digit code which indicates the spectroscopic database used for CASSIS line identification and modelling.}  
\label{tbl-2}
\begin{tabular}{llrrrrcclll}
\hline
Species  & TAG &  Transition  & Frequency & $E_u$ & $A_{ul}$ & HPBW & $\eta_{mb}$   & $\int T_{A}^*dv$  & FWHM & $V_{lsr}$   \\
             &         &                    & MHz &  K & $10^{-5}$ s$^{-1}$  & $^{\prime\prime}$ &  &   K km s$^{-1}$  & km s$^{-1}$ & km s$^{-1}$ \\
\hline
HCN          &  27501 &  1 -- 0  & 88631.602	&	4.3	    &	2.4	&	27.8	&	0.85	&	3.40(.01)  &	1.4(.1)	   & 2.2 (.1)  \\
                 &             & 3 -- 2	 &	265886.434	&	25.5	    &	83.6	&	9.3	&	0.59	&	1.86(.20)  &	0.8(.1)     & 2.2 (.1) \\
DCN          & 28509  & 1 -- 0   & 72414.905  &  3.5      & 1.3    & 34.0 & 0.86  &  0.49(.02) & 1.5(.2)     &  2.4(.1) \\
                 &             & 2 -- 1   &144828.001 & 10.4     & 12.7 & 17.0  & 0.79  &  0.32(.01) & 1.0(.1)     & 2.5(.1)    \\
                 &             & 3 -- 2    & 217238.538 & 20.9    & 46.0 & 11.3  &  0.65 &  0.10(.01) & 0.96(.04) & 2.8(.1)  \\
H$^{13}$CN& 28501& 1 -- 0	 &	 86340.184	 &	 4.1	    &	2.2  & 28.5	&	0.85	&	0.27 (.02) & 1.3(.1)     & 2.7(.1)  \\
                  &            & 2 -- 1	 & 172677.851	 & 12.4	    &21.4  &14.3	&	0.75	&	0.10(.02)  &	1.4(.3)     & 2.4(.1) \\
                  &            & 3 -- 2	 & 259011.798	 & 24.9     & 77.3 &	9.5	&	0.60	&	0.08(.01)	&	1.3(.1)     & 2.7(.1) \\
HC$^{15}$N & 28506&1 -- 0   & 86054.961   &   4.1     & 2.2   &  28.6 & 0.86 &  0.09(.02) & 1.5(.2)     & 2.6(.1) \\
                  &             & 2 -- 1  & 172107.956 &  12.4    & 21.1 &  14.3 & 0.75 & 0.06(.01)  & 1.8(.4)     & 2.0(.2) \\ 
                  &             & 3 -- 2  & 258157.100 &  24.8    & 76.5 &   9.5  & 0.60 & 0.04(.01)  & 1.5(.2)     & 2.6(.1) \\  \hline
HNC          & 27502   & 1 -- 0	 & 90663.568	 &	 4.4 	     &	2.7	&	27.1 &	0.85 &	  3.1(.1)  &	1.8(.1)     &2.4(.1)  	\\
                 &              & 3 -- 2	 & 271981.142	 &  26.1    &93.4  &	9.0 	&	0.58 &	  1.3(.1)  	&	0.99(.1)   & 2.5(.1)	\\
DNC          & 28508   & 1 -- 0  & 76305.701  &  3.7      & 1.60  & 32.2 & 0.86 & 0.90(.01)  & 1.7(.1)       & 2.7(.1) \\
                &               & 2 -- 1  &152609.746 & 11.0     & 15.4  & 16.1 &  0.77& 0.69(.01)  & 1.0(.1)       & 2.6(.1)  \\
                &               & 3 -- 2  &228910.489 & 22.0     & 55.7  & 10.7 &  0.64& 0.26(.01)  & 0.8(.1)      & 2.7(.1)\\
HN$^{13}$C& 28515 & 1 -- 0  &	87090.825   &	4.2	    &	2.4	&	28.3 &	0.85& 0.29(.01)  & 1.4(.1)      &2.4 (.1) 	\\
                   &            & 3 -- 2  &261263.310 & 25.1     & 82.8  &   9.4  & 0.60& 0.079(.004)&0.8(.1)    & 2.3(.1)\\
H$^{15}$NC& 28006 & 1 -- 0  & 88865.715  & 4.3       & 2.5    & 27.8  & 0.85& 0.063(.004)&1.4(.1)    & 2.7(.1)  \\
                   &            & 3 -- 2  &266587.800 &25.6      &87.9   & 9.2   & 0.59 & 0.026(.005)&1.5(.4)    & 2.9(.2) \\ 
\hline
\end{tabular}
}
\end{table*}

\section{Results}
\label{results}

\subsection{Spatial distribution}
In Fig.~\ref{map-hcn}, we show in white contours the spatial distribution  of the HCN $J$=3--2 velocity-integrated emission between $-20$ and $+5\kms$ over the southern outflow lobe. The emission 
draws two extended features of different Parallatic Angles, which coincide with the two CO outflow cavities identified by \citet{Gueth1996}. The emission peaks at the nominal position of the outflow shock position B1. Secondary emission peaks are  found towards the shock positions B0 and B2, at  about $30\arcsec$ north and south of B1, respectively \citep{Bachiller2001}. Hardly any emission is detected along the outflow between B0 and the protostar (located North, red cross in Fig.~1).
The systemic velocity of L1157 is $+2.7\kms$ \citep{Bachiller1997}. We show in color scale in Fig.~\ref{map-hcn}  the HCN $J$=3--2 emission integrated over the three velocity intervals (in $\kms$): [$-$10.2; $-$2.7], [$-2.7$; $+$4.3] and [$+$4.3; $+$6.3], from left to right.   We find that the low-velocity emission (middle panel in Fig.~\ref{map-hcn}) is associated with the bright shocked regions B0, B1, and B2, as well as faint and extended emission tracing the whole outflow lobe. As shown by \citet{Podio2016}, the outflow is oriented almost in the plane of the sky 
(inclination $i\sim73\degr$). We propose that the HCN emission between $-2.7$ and $+4.3\kms$ could arise from low-velocity shocks associated with the cavity walls of the outflow, which would explain the extended emission at blue- and red-shifted velocities. 
This is consistent with the distribution of the HCN $J$=1--0 emission line as mapped by  \citet{Benedettini2007}  at $\approx 5\arcsec$ with the Plateau de Bure Interferometer. 
The high-velocity blueshifted gas traces the young shocks B0 and B1 (left panel in Fig.~\ref{map-hcn}), while the redshifted emission is only detected toward the oldest B2 shock (right panel). These maps indicate that the shocked gas in B0 and B1 emit at a velocity distinct from that of B2. Interestingly, the brightest HCN $J$=3--2 emission coincides with the apex of the outflow cavities, consistent with the high angular resolution NOEMA images of H$^{13}$CN and HC$^{15}$N \citep{Busquet2017,Benedettini2021}.

To summarize, HCN emission is detected throughout most of  the southern outflow lobe, from B0 to the tip of the outflow at position B2, and the brightest emission regions are associated with outflow shocks.  This is consistent with the distribution of the HCN $J$=1--0 emission line as mapped by  \citet{Benedettini2007}  at $\approx 5\arcsec$ with the Plateau de Bure Interferometer.

\subsection{L1157-B1} 

All the transitions of the HCN isotopologues and most of the transitions of the HNC isotopologues present  in the spectral range of ASAI, between 72 and 272 GHz, were detected towards the shock region (see Table~1).  Due to  instrumental (spectral bandwidth) and observational (rms noise)  limitations, only the ground state transitions of HN$^{13}$C and  H$^{15}$NC were detected with 
a SNR $>$ 3.  The  profiles of the transitions display broad linewidths (up to $\approx 10\kms$). 
All the line profiles peak at $V_{lsr}\approx 0\kms$ and trace blueshifted gas, in agreement with the other molecular tracers reported in previous works \citep{Lefloch2012, Codella2012, Gomez2015}.

The excellent rms of the data permits detection of emission up to $V_{\rm LSR}\approx$ $-20\kms$ in the  different  transitions of HCN. 
Several transitions of DCN, H$^{13}$CN and HC$^{15}$N were also detected towards L1157-B1 and their  profiles display similar broad linewidths, confirming the shock  association (Fig.~\ref{l1157b1-hcn}). The detection of bright deuterated emission suggests that despite shock processing, the gas has preserved part of the initial chemical conditions, an effect already pointed out by \citet{Codella2012} and \citet{Fontani2014}, who reported the detection of singly deuterated isotopologues of H$_2$CO, CH$_3$OH, NH$_3$. A small systematic velocity shift (less than $0.5\kms$) is measured  between the emission peaks of the HCN and HNC line profiles, suggesting they are tracing different regions inside the shock. 

We note that the HNC isotopologues (HN$^{12}$C, HN$^{13}$C, H$^{15}$NC) display much weaker lines intensities than the HCN isotopologues. Line intensities and fluxes of the ground state transitions  are weaker  by  one order of magnitude. There is however one exception, which is  the case of deuterated isotopologues DNC and DCN, whose peak  intensities are similar within  a factor of 2. 

\begin{figure}
\centering{
\includegraphics[width=0.9\columnwidth,keepaspectratio]{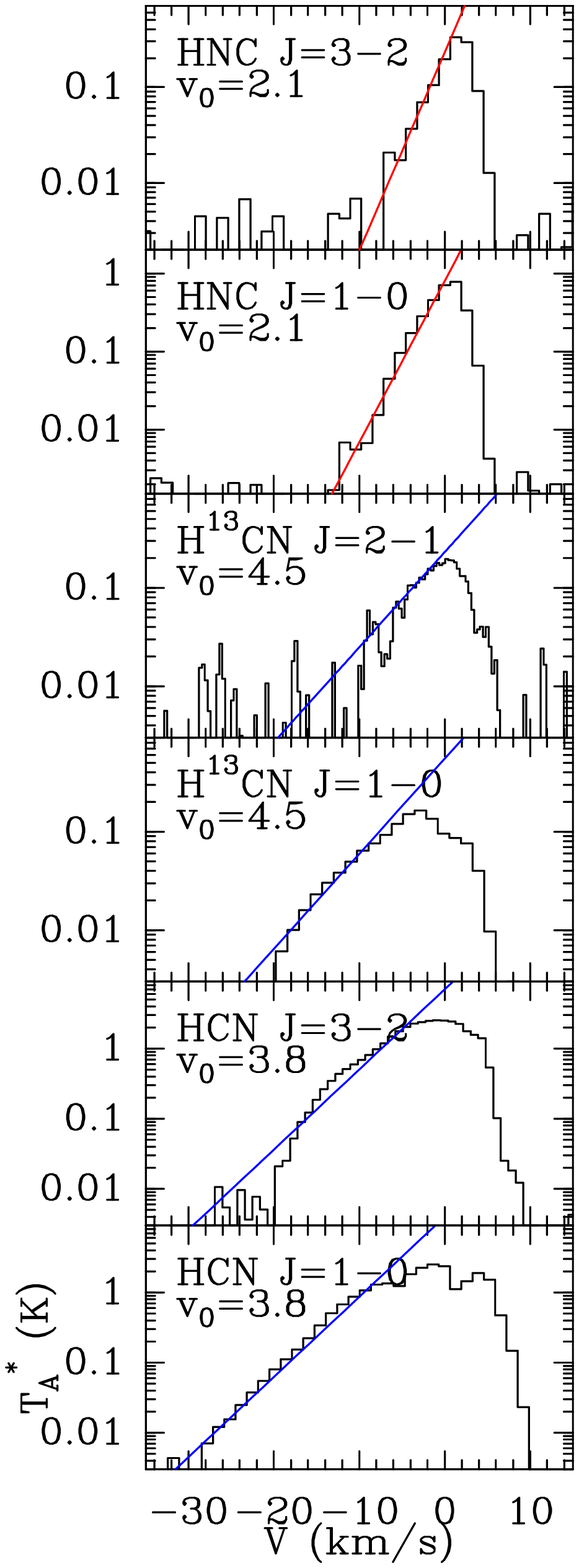}
\caption{Montage of emission line profiles  observed towards L1157-B1 displayed in linear-logarithmic scale, from top to bottom:
HNC $J=3-2$, $J=1-0$, H$^{13}$CN $J=2-1$, $J=1-0$, HCN $J=3-2$, $J=1-0$.
In each panel, we have superposed in red (blue) a fit to the spectral slope of the type $T(v)\propto \exp(-V/V_0)$ with $V_0\simeq$ 2.0 (4.0)$\kms$, associated with the component $g_3$ ($g_2$) of the outflow. The exponent $v_0$ is given for each transitions in the corresponding panel.}
\label{profiles}}
\end{figure}

Following the approach presented in \citet{Lefloch2012}, we have fitted the line profiles by an exponential law of the  type $\exp(-V/V_0)$, a procedure which  is obviously limited by the SNR of the data. In practice, we considered the 
lower-$J$ transitions of HCN, H$^{13}$CN and HNC. The results are displayed in Fig.~\ref{profiles}, where the spectra are displayed in a linear-logarithmic scale. 
The spectral slope of the HCN (H$^{13}$CN)  line profiles considered can be fitted by an exponential law of the type $\exp(-V/V_0)$,
with $V_0$= $3.8\kms$ ($V_0$= $4.5\kms$; Fig~\ref{profiles}). The similar values obtained for $V_0$ indicate that we are indeed probing the same gas in the different transitions of the same molecular species. A small deviation with respect to the fit is observed at $V < -10\kms$ in the HCN $J$=1--0 line profile. This is probably due to the overlap between the contributions of the different hyperfine satellites. Another possibility would be the presence of spatial gradients of emission across the telescope beam \citep{Benedettini2013}. 
 
The spectral slope $V_0$  of the HCN line profiles is very similar to that found for the CO emission from the L1157-B1 outflow cavity (component $g_2$) by \citep{Lefloch2012}, as indicated by the similarity of the fits and the derived exponent values ($V_0=4.4\kms$).  This implies that in the velocity range between $-20$ and $-5\kms$, both species are probing the same gas component. This is consistent with the  analysis of the HCN $J$=3--2 spatial distribution obtained with the IRAM 30m telescope at $9.3\arcsec$ (Sect.~4.1) and the HCN $J$=1--0 and H$^{13}$CN 
$J$=2--1 with the IRAM interferometer at a few arcsec resolution \citep{Benedettini2007,Busquet2017}. As can be seen in Fig.~\ref{profiles}, the HNC line profiles look very different from those of HCN in a linear-logarithmic scale. The velocity range of the emission is narrower and  stops at $-10\kms$, whereas the emission of HCN extends up to $-30\kms$. Quantitatively, the  HNC line profiles  are well fitted by an exponential function $\exp(-V/V_0)$ with $V_0$= $2.1\kms$, i.e. a value of $V_0$ half the value obtained for HCN.  This supports the idea that HNC and HCN are tracing different regions. We note that the $V_0$ value  obtained for HNC is actually  similar to that found for the CO  component $g_3$ ($V_0=2.5\kms$), which \citet{Lefloch2012} showed to be associated with the L1157-B2 outflow cavity.

In L1157-B1, the HCN  line profiles  follow an exponential signature  
$\exp(-V/V_0)$, with the same $V_0$ value for the rotational transitions $J$= 1--0 and $J$= 3--2, so that the HCN  line intensity ratio  $J$= 3--2/$J$=1--0 is almost constant for velocities beyond $-6\kms$ (see Fig.~\ref{profiles}).  Hence,  the HCN excitation conditions are essentially independent of velocity, at least in this velocity range. Similar results were previously reported  for CO \citep{Lefloch2012}.
The similarity of the spectral slopes of HCN and CO between $-20$ and $-5\kms$ implies that the HCN/CO line flux ratio is constant, and since the excitation conditions of both species are independent of velocity in this interval, it means that the relative abundance ratio is also independent of velocity in this velocity range. 

The HCN/H$^{13}$CN total line flux ratio is  $\approx 19$ and 42 for the $J$=1--0 and $J$=3--2 transitions, respectively. Assuming similar excitation conditions for both isotopologues and an elemental abundance ratio $^{12}$C/$^{13}$C equal to 66 (see below), close to the local ISM value \citep{Milam2005},  we then derive line opacities $\tau_{10}\simeq 3$ and $\tau_{32}\simeq 0.8$
for the $J$=1--0 and $J$=3--2, respectively, i.e. the gas is moderately optically thick. The HNC/HN$^{13}$C takes lower values, $\approx 11$ and $16$ for the $J$=1--0 and $J$=3--2 transitions, respectively, which implies opacities of about 5 and 4 for the main isotopologue, respectively, still under the assumption of a standard elemental abundance ratio. 


\subsection{L1157-mm}

Emission from the  HCN and HNC isotopologues was detected towards the protostellar envelope of L1157-mm. All the detected molecular transitions display narrow linewidths between 0.8 and $1.8\kms$, as expected for cold and quiescent protostellar gas (Figs.~\ref{l1157mm-hcn}--\ref{l1157mm-hnc}).  This permits identification of the hyperfine satellites, whose  emission was hidden otherwise in the outflowing gas at position B1. The signature of the protostellar outflow is detected in the HCN and HNC $J$=1--0 lines, as a broad, low intensity component with a typical linewidth of $5\kms$. 

Overall, comparison of the  HCN and HNC isotopologues  emission shows  that the transitions from same rotational levels  display  similar similar intensities for each pair of isomers, as can be seen in Figs.~\ref{l1157mm-hcn}-\ref{l1157mm-hnc}  (see also Table~\ref{ratios}). Since the corresponding  line intensity ratios are only weakly dependent on the excitation temperature, this suggests that both isomers and their rare isotopologues should have similar abundances. Note that the situation differs very much from L1157-B1, where much brighter  emission  is detected in the HCN isotopologues. 

The HCN/H$^{13}$CN (HNC/HN$^{13}$C) line flux ratio  is 11 and 16 (12 and 23) for the $J$=1--0 and 3--2 transitions, respectively. Adopting a standard elemental abundance ratio $^{12}$C/$^{13}$C = 66, we then derive HCN line optical depths 
$\tau_{10}\simeq$ 5.5 
and $\tau_{32}\simeq$ 2.3, for the  $J$=1--0 and $J$=3--2 transitions, respectively.  Similar results are obtained for HNC. Hence, both the HCN and HNC emissions are optically thick.  
Therefore, we have used the $^{13}$C isotopologues to determine the total gas column density of HCN and HNC. 

To summarize, a simple comparison of the  line intensities of the  HCN and HNC isotopologues measured towards L1157-mm suggests that both isomers are present in the pre-shock phase with similar abundances. 
The pre-shock gas displays hints of isotopic fractionation enrichment, in particular deuteration, which is  partially preserved  through  the shock.  
A strong differentiation is observed in the shock between HCN and HNC with the emission of the former strongly enhanced over the latter. Analysis of the spectral signature suggests that both isomers are  tracing different regions  in the  shock region.

\subsection{Line excitation}

The above analysis of the  line profiles shows hints of chemical evolution in the passage of the shock. 
In order to gain more insight on the shock impact on nitrile chemistry, we have first determined the excitation conditions of the different molecular species (excitation  temperature, column density). The detection of the rare isotopologues  offers an opportunity to constrain directly the evolution of isotopic fractionation through the passage of a protostellar shock. 
In the case of L1157-mm, rather than a detailed physical and chemical description of the structure of the envelope, our goal here is to determine the relative abundances of these various species, which we take as representative of the chemical composition of the pre-shock gas in the L1157-B1 region. 

In order to follow a systematic approach in  the molecular abundance determination, we decided to model the line profiles in the approximation of Local Thermodynamical Equilibrium (LTE) using CASSIS\footnote{http://cassis.irap.omp.eu/} \citep{Vastel2015}. In the case only one transition was detected (HC$^{15}$N, HN$^{13}$C in L1157-B1), an excitation temperature of $7\K$ was adopted, in good agreement with the other rare isotopologues. Line profiles towards L1157-B1 could usually be reproduced with one single Gaussian component, of Full Width at Half Power (FWHP) consistent with our determination of the line parameters (see Table~\ref{tbl-1}). In order to reproduce as accurately as possible the  observed line profiles, we made use of the CASSIS private database which takes into account the hyperfine spectroscopic properties of HCN isotopologues.  Towards L1157-mm, it was necessary to introduce a second component of broad linewidth ($5\kms$) and low intensity in order to reproduce the outflow signature (see Sect.~4.3).

We have adopted a size of $60\arcsec$ for the envelope of L1157-mm,  a  value in good agreement with the size of the envelope 
observed in N$_2$H$^+$ \citep{Tobin2013}. In practice, this value allowed us to best simultaneously reproduce the emission of the various transitions of a given molecular species under the assumption of one single excitation temperature. 
We have adopted a total gas column density of $2\times 10^{22}\cmmd$, based on the  \htwo\ column density maps of  the region obtained in the Herschel Gould Belt Survey (HGBS) of nearby star-forming molecular clouds \citep{DiFrancesco2020}. We note that this \htwo\ column density is about a factor of 3 higher 
than the determination obtained by \citet{Mendoza2018}. The latter value ($6\times 10^{21}\cmmd$) was derived from the  $^{13}$CO $J$=1--0 line in the ASAI spectrum of the protostar, adopting a canonical abundance of $1.6\times 10^{-6}$.  As a matter of fact, the low gas kinetic temperature ($\sim 10\K$) and  the relatively high gas density ($\sim $ a few $10^4\cmmt$) of the envelope are such that CO is most likely depleted at large scale in the envelope by a factor of a few, as is commonly measured in cold cores \citep[see e.g.,][]{Bacmann2002}. Both column density values can be reconciled if CO is moderately depleted by a factor of about 3--4 over the envelope.  We have adopted a linewidth of $\sim 1\kms$ for the line profile modelling in L1157-mm, in agreement with the results of our gaussian fits to  the line spectra (see Figs.~\ref{l1157mm-hcn}--\ref{l1157mm-hnc} and Table~2).

In the case of L1157-B1, previous observational work showed that the molecular gas emission arises from a region with a typical gaussian size  (FWHM) in the range $20\arcsec$ -- $25\arcsec$ \citep{Gueth1996,  Lefloch2012,  Benedettini2013}. This is consistent with the distribution of the HCN $J$=3--2 emission (Fig.~\ref{map-hcn}), which appears extended  between the protostar and the B1 shock position.  The high-$J$ transitions of the HCN and HNC isotopologues  lie in a frequency range for which the main-beam of the telescope is comparable to or smaller than the size of the shock region, as estimated from interferometric observations (see Tables~\ref{tbl-1}--\ref{tbl-2}).  For these transitions, the emission can be considered as extended and  the main-beam temperature becomes a satisfying approximation to the molecular line brightness temperature. 
The case of the ground state transitions of the HCN and HNC isotopologues is more complicated for two reasons: first, the telescope beamwidth (HPBW) is comparable to the molecular emission size; second, the typical physical conditions 
($n(\htwo)\sim 10^5\cmmt$; $T_{kin}$= $60\K$--$90\K$) are such than the rotational transitions present in the millimeter bands are subthermally excited. The $J$=1--0 excitation temperature is higher than that of the 
 $J$=3--2 line, as can be checked by a simple calculation using a radiative transfer code in the Large Velocity Gradient approximation, like Madex (Cernicharo et al. 2012). In the LTE approximation, the $T_{ex}$ of the $J$=1--0 line is underestimated by the fitting procedure, and the approximation is no longer very satisfying. However, the approximation of extended, or main-beam averaged, emission  compensates for the previous effect, which mainly affects the ground state transitions, as explained above.  We found that this approximation yields much better quality LTE fits. The source-averaged column density is then obtained by correcting the fitted (main-beam averaged) column density for the main-beam filling factor. 

The results of our LTE analysis (source-averaged column density $N$, excitation temperature $T_{ex}$) and the corresponding molecular abundances are summarised in Table~\ref{LTEfitting}.
Due to the uncertainties in the derivation of the HCN column densities from the  $^{12}$C isotopologue, we have reported for HCN and HNC  the values obtained from the rare $^{13}$C isotopologues, adopting an elemental abundance ratio $^{12}$C/$^{13}$C = 66 (see below Sect.~5.1). They are indicated with a (*) in Table~\ref{LTEfitting}.  

The best fits to the individual line profiles obtained with CASSIS are superimposed  in red on the spectra in Figs.~\ref{l1157b1-hcn}--\ref{l1157mm-hnc}. 
The error bars on $T_{\mathrm{rot}}$ and $N$ were estimated from LTE fits to the line profiles taking into account the absolute flux calibration uncertainties (see Sect.~3.3) and the statistical rms noise. 
Our column density determinations for HCN and HNC  towards L1157-mm and L1157-B1 are in good agreement with those of \citet{Bachiller1997}. 
Recently, \citet{Benedettini2021} observed the H$^{13}$CN and HC$^{15}$N $J$=1--0 line emission at $4\arcsec$ resolution toward L1157-B1 using NOEMA, as part 
of the Large Program SOLIS. Their  radiative transfer modelling of both lines yielded $N$(H$^{13}$CN) $\sim 7\times10^{12}\cmmd$ and $N$(HC$^{15}$N) $\sim2 \times10^{12}\cmmd$. Our observational determinations are in satisfying agreement with these values (see Table~\ref{LTEfitting}).

\section{Discussion}

In this section, we discuss the properties of the HCN and HNC  isotopologues in the pre-shock gas, based on the L1157-mm envelope properties, and their evolution across the shock in L1157-B1. The molecular abundances and line ratios of interest derived towards L1157-B1 and L1157-mm are presented in Table~\ref{ratios}. Using the time dependent gas-grain chemical and parametrised shock model UCLCHEM \citep{Holdship2017} we simulate the pre-shock abundances for the gas and the solid phase, before the arrival of the shock, as well as during and after the shock has passed. We discuss the modelling findings with an emphasis of the behaviour of HCN and HNC.

\label{physical}
\begin{table*}
\begin{center}
\caption{Excitation conditions (source-averaged column density $N$, rotational  temperature $T_{\rm rot}$, abundance $X$) of the HCN and HNC isotopologues detected towards L1157-B1 and L1157-mm.  Emission sizes of $60\arcsec$ and $20\arcsec$ were adopted for  L1157-mm and L1157-B1, respectively.  A total \htwo\ column density of $2\times 10^{22}\cmmd$ and $2\times 10^{21}\cmmd$ was adopted for L1157-mm and L1157-B1, respectively \citep{Lefloch2012,DiFrancesco2020}, in order to compute the molecular abundances. 
We follow the convention $a(b)= a \times 10^b$. The HCN and HNC parameters are marked  with a * to indicate that they were obtained from the $^{13}$C isotopologue, adopting an abundance ratio $^{12}$C/$^{13}$C= 66 (see Sect.~5.1). }
\label{LTEfitting}
\begin{tabular}{lccccccc} \hline
                     & \multicolumn{3}{c}{L1157-mm} &  & \multicolumn{3}{c}{L1157-B1}  \\
                           \cline{2-4}   \cline{6-8}\\
Species           & $T_{\mathrm{rot}}$   &  $N$         & $X$            &    &  $T_{\mathrm{rot}}$ & $N$           &    $X$            \\
                      & ($\K$)   & ($\cmmd$)&                  &    &  ($\K$)    & ($\cmmd$)&                \\       \hline
Main Body \\                    
HCN(*)            &  6.7$^{+0.7}_{-0.5}$  &  4.6$^{+0.6}_{-0.6}$(13)& 2.3$^{+0.3}_{-0.3}$(-9)   &     & 7.2$^{+0.6}_{-1.0}$   &  6.6$^{+1.1}_{-0.6}$(14) & 3.3$^{+0.6}_{-0.3}$(-7)   \\    
                       &                                  &                                        &                                         &     &           &              &                        \\ 
H$^{13}$CN   & 6.7$^{+0.7}_{-0.5}$   &  7.0$^{+1.0}_{-1.0}$(11) &  3.5$^{+0.5}_{-0.5}$(-11) &    &  7.2$^{+0.6}_{-1.0}$ & 1.0$^{+0.4}_{-0.1}$(13) & 5.0$^{+2.0}_{-1.0}$(-9)   \\
                       &                                  &                                        &                                         &     &               &               &                       \\
DCN                &  4.9$^{+0.2}_{-0.2}$ &  2.8$^{+0.5}_{-0.3}$(12) &  1.4$^{+0.3}_{-0.2}$(-10)&     &  6.5$^{+0.5}_{-0.5}$&  2.1$^{+0.5}_{-0.5}$(12)  & 1.1$^{+0.2}_{-0.3}$(-9)  \\
                      &               &                 &                   &     &                                  &                                         &                    \\
HC$^{15}$N   & 7.7$^{+0.5}_{-0.4}$ &  1.2$^{+0.3}_{-0.2}$(11)   &  0.7$^{+0.1}_{-0.1}$(-11)&    &6.8$^{+0.4}_{-0.5}$&  2.0$^{+0.5}_{-0.3}$(12)  & 1.0$^{+0.3}_{-0.2}$(-9)  \\ 
                      &               &                 &                   &     &                                  &                                        &                     \\
\hline
HNC(*)            &  6.8$^{+0.3}_{-0.3}$&  4.2$^{+0.8}_{-0.5}$(13)   &  2.1$^{+0.4}_{-0.3}$(-9)     &   &  7.0$^{+1.3}_{-1.5}$ & 3.5$^{+0.3}_{-0.6}$(13)  & 1.7$^{+0.2}_{-0.2}$(-8)    \\
                      &               &                 &                   &     &                                  &                                        &                      \\
HN$^{13}$C   &  6.8$^{+0.3}_{-0.3}$ & 6.3$^{+1.2}_{-0.7}$(11)   &  3.2$^{+0.6}_{-0.4}$(-11)    &   &7.0$^{+3.0}_{-1.7}$  &  5.3$^{+0.6}_{-0.6}$(11)  & 2.6$^{+0.3}_{-0.2}$(-10)   \\
                      &               &                 &                   &     &                                  &                                        &                      \\
DNC               &  5.7$^{+0.3}_{-0.2}$ & 2.5$^{+0.5}_{-0.3}$(12)   &  1.3$^{+0.2}_{-0.2}$(-10)    &    &6.0$^{+2.0}_{-2.0}$ &  3.2$^{+1.8}_{-1.4}$(11)  & 1.6$^{+0.9}_{-0.7}$(-10)  \\
                      &               &                 &                   &     &                                  &                                         &                    \\
H$^{15}$NC   &  7.7$^{+0.5}_{-0.6}$ & 1.3$^{+0.3}_{-0.4}$(11)  &  0.7$^{+0.1}_{-0.3}$(-11)    &   &7.0$^{+3.0}_{-1.5}$  & 1.8$^{+0.7}_{-0.6}$(11)   & 8.5$^{+3.5}_{-2.5}$(-11)  \\ 
\hline
Outflow component \\ 
HCN              &   5.0      &   4.0(12)     &     2.4(-8)          &            &      -       &        -    & -      \\  
HNC              &  5.0       & 1.5(12)       &     0.9(-8)          &           &      -        &       -     & -       \\
\hline
\end{tabular}
\end{center}
\end{table*}

\begin{table}
\begin{center}
\caption[]{Isomer and isotopic abundance ratios of HCN and HNC in L1157-mm and L1157-B1. For each source, the ratios  are computed adopting a canonical elemental $^{12}$C/$^{13}$C abundance ratio  of  66 (see also Sect.~5.1). }
\label{ratios}
\begin{tabular}{lrrr} \hline
                                          & L1157-mm &  & L1157-B1  \\
\hline
HCN/H                                 &1.2$^{+0.2}_{-0.2}$(-9) &  & 1.7$^{+0.4}_{-0.2}$(-7)               \\
HCN/H$^{13}$CN                 &    66                        &  &      66                  \\
HCN/HC$^{15}$N                 &    $383\pm145$     & &    $330\pm 110$    \\ 
HCN/DCN                             &    $16 \pm 5$         &  &   $314\pm 110$     \\
H$^{13}$CN/HC$^{15}$N     &    $5.8\pm 2.2$      &  &   $5.0\pm 2.5$       \\ \hline
HNC/HN$^{13}$C                 &     66                      & &    66                     \\ 
HNC/H$^{15}$NC                 &   $323\pm 160$    & &   $194\pm 100$          \\    
HNC/DNC                             &   $17 \pm 7$         &  &  $109\pm 70$            \\
HN$^{13}$C/H$^{15}$NC     &   $4.8\pm 2.4$      & &   $3.0\pm 1.5$          \\  \hline
HCN/HNC                             &   $1.1\pm 0.4$      & &   $19\pm 6$             \\  
H$^{13}$CN/HN$^{13}$C     &   $ 1.1\pm 0.4$     & &   $19\pm 8$              \\
HC$^{15}$N/H$^{15}$NC     &   $0.9\pm 0.5$      &  &    $11\pm 6$               \\
DCN/DNC                             &   $1.1\pm 0.4$      &   &  $6\pm 5$           \\ \hline
\end{tabular}
\end{center}
\end{table}

\subsection{$^{12}$C/$^{13}$C elemental abundance ratio}

We have used the detected transitions of HC$_3$N and its rare $^{13}$C isotopologues observed towards L1157-B1 to estimate  the elemental abundance ratio $^{12}$C/$^{13}$C in the L1157 star-forming region. The  transitions of the rare isotopologues $J$=10--9, $J$=11--10, $J$=12--11 of H$^{13}$CCCN, HC$^{13}$CCN and HCC$^{13}$CN were presented in \citet{Mendoza2018}. The transitions of same quantum numbers of the different $^{13}$C isotopologues are very close in frequency, separated by a few GHz at most. These different transitions were observed simultaneously in the ASAI survey, with the advantage of minimizing the calibration uncertainty, which is then dominated by the rms noise of the spectrum.  
\citet{Mendoza2018} showed that the emission of all the HC$_3$N isotopologues detected in the millimeter range is optically thin. Hence, the line flux ratio of the $^{12}$C to the sum of the $^{13}$C isotopologues is equal to three times the elemental abundance ratio $^{12}$C/$^{13}$C. For the $J$=10--9 and 
$J$=12--11 transitions, we found  $^{12}$C/$^{13}$C= $66\pm6$. For the $J$=11-10, we discarded the H$^{13}$CCCN line whose intensity is almost twice as bright as the lines of HC$^{13}$CCN and HCC$^{13}$CN, a dissymmetry which is observed only in this transition and not in the $J$=10--9 and 
$J$=12--11. It then comes $^{12}$C/$^{13}$C= $70\pm 10$, in agreement with the determinations from the $J$=10--9 and 
$J$=12--11. 
This direct observational determination of the $^{12}$C/$^{13}$C ratio in the L1157 star forming region is actually very close to the elemental abundance ratio in the local ISM, equal to 68, \citep{Lucas1998,Milam2005} and the value assumed in our previous works (see Sect. ~2).

Both the  HCN/H$^{13}$CN and HNC/HN$^{13}$C column density ratios display values well below the canonical $^{12}$C/$^{13}$C elemental abundance ratio in the ISM, equal to 68  \citep{Lucas1998,Milam2005}. In particular, a low $^{12}$C/$^{13}$C ratio of $\approx 11\pm 4$ is measured towards towards the cold protostellar core L1157-mm. We first note that models of $^{13}$C isotopic fractionation in cold cores lead to an opposite effect, i.e. a H$^{12}$CN/H$^{13}$CN (HN$^{12}$C/HN$^{13}$C) ratio higher than the $^{12}$C/$^{13}$C elemental abundance ratio \citep{Langer1984,Roueff2015}. 
 A similar effect was reported by citet{Daniel2013} in their  detailed study of nitrogen hydrides and nitriles in the Class 0 protostellar core  Barnard 1. The authors concluded that the apparently low H$^{12}$CN/H$^{13}$CN ratio was most likely a bias caused by the simple hypothesis made in the radiative transfer modelling of the line excitation and the geometry of the astrophysical source. In particular, the excitation of the  HCN and HNC  ground-state transitions also depends on the extended, low-density gas surrounding the dense core, so that a detailed multi-dimensional radiative  transfer model of the source taking into account the full density and velocity fields is needed in order to properly reproduce the line profiles .

From what precedes, we conclude that there  is no convincing evidence for $^{12}$C/$^{13}$C fractionation of HCN/HNC in the star-forming region. Because of the high optical depth of the HCN and HNC lines, which cast some uncertainties on the determation of the  HCN and HNC total column densities, these were obtained  from the column densities of their  optically thin $^{13}$C isotopologues and taking into account our  estimate of the elemental abundance  $^{12}$C/$^{13}$C= $66\pm 6$. 

\subsection{Isomers}

In the pre-shock gas, we find similar molecular abundances  of HCN and HNC, close to  $2\times 10^{-9}$, and a a relative abundance ratio 
HCN/HNC= H$^{13}$CN/HN$^{13}$C= $1.1\pm 0.4$.  
These values were derived from a direct to the L1157-mm line profiles with CASSIS, taking into account the line optical depths. 
Similar results were obtained when considering the rare isotopologues ratios DCN/DNC (= $1.1\pm 0.4$) and HC$^{15}$N/H$^{15}$NC 
(= $0.9\pm 0.5$). The consistency between these values determined independently, makes us confident about the reliability of our method. 
The molecular abundances of HCN and HNC are similar to those reported in previous surveys  of dark cloud cores and low-mass star forming cores \citep[e.g.,][]{Hirota1998,hily2010,Daniel2013}, suggesting that the envelope of L1157-mm has kept memory of the prestellar phase. \citet{Hirota1998} pointed out  that since no difference is observed between the values measured in prestellar and protostellar cores, the evaporation of HCN and HNC from dust grains does not contribute significantly to the observed emission in the cold gas.

Comparison of the molecular abundances of L1157-mm (pre-shock gas) and the shocked region L1157-B1 (see Table~\ref{LTEfitting}) shows that the  HCN and HNC isomers behave differently across the shock.  First, we  observe an increase in the abundance of all the HCN isotopologues. The rare H$^{13}$CN and HC$^{15}$N (and HCN) abundances increase by a similar factor of 140 while the DCN abundance increases much less, by a factor  $\approx 8$. 
Unlike HCN,  the abundances of all the HNC isotopologues vary by a much smaller factor of a few ($\approx$ 8--12). The abundance DNC remains almost unchanged. Overall, the abundances of the HNC isotopologues seem barely affected by the passage of the shock. 
From a ratio HCN/HNC with an initial value of 1 in the quiescent gas, it is now  $19\pm 6$ in the post-shocked gas. The isomer ratio displays values in the range 11--19  which are consistent, within the uncertainties, for H$^{13}$CN/HN$^{13}$C and HC$^{15}$N/H$^{15}$NC.  The deuterated isomers seem to behave differently with a much lower 
increase of the  DCN/DNC ratio. 

In the protostellar outflow near L1157-mm, the high sensitivity of the ASAI data has permitted detection of the emission of  the ground state transition of HCN and HNC. This emission can be modelled as a broad component of $\approx 5\kms$ and low excitation (Table~\ref{LTEfitting}). Interestingly, the HCN/HNC abundance (column density) ratio is close to 3, suggesting that the relative abundance of HCN with respect to HNC is enhanced in the formation of the outflow. 

\subsection{Isotopic fractionation}

\subsubsection{Deuterium}
It is well established that strong Deuterium enrichment takes place in molecular material of cold, dark and prestellar cores \citep{Caselli2012}. The first evidence of molecular deuteration in L1157-mm were brought by \citet{Bachiller1997}, with the detection of the deuterated forms of HCN and HCO$^+$, and  a molecular  D/H ratio of  0.018 in the cold envelope.
Our own determinations, which are based on the modelling of a larger number of rotational transitions, yield a somewhat higher D/H ratio 
$0.06\pm 0.02$ for both HCN and HNC,  in rough agreement with \citet{Bachiller1997}. 

At the passage of the shock,  the DCN abundance increases by a factor 7  whereas DNC remains unchanged. Overall, the magnitude of these variations remain small in front of those affecting the main isotopologues and indicate that, at first order, the abundances of these deuterated species are only moderately affected by the passage of the shock, with a different behaviour for DCN and DNC. 
As for L1157-B1, previous estimates of the DCN/HCN ratio were obtained by \citet{Codella2012}, based on a reduced number of transitions of  H$^{13}$CN and DCN. Our present determination of $(3.1\pm 1.0)\times 10^{-3}$ is consistent with  their previous estimate D/H = (0.5--3)$\times 10^{-3}$. More recently, \citet{Busquet2017} reported similar deuterium fractionation ratios ($\approx 4.0\times 10^{-3}$) in the small-scale structures detected in the bow-shock, with no significant spatial variation. 
Our unbiased analysis of the deuterated isomers confirms that  the molecular D/H ratio decreases across the shock and it allows us to quantify the magnitude of the effect: a factor 26 and 8 for HCN and HNC, respectively. 

The behaviour of DCN/HCN in the B1 shock position has already been studied in detail by \citet{Busquet2017} based on high-angular resolution observations of H$^{13}$CN $J$=2--1 and DCN $J$=2--1 combined with the UCLCHEM code and adopting the deuterated network of \citet{Esplugues2013}. The results from the model show that the DCN/HCN abundance ratio varies with the passage of the shock. The observed DCN/HCN ratio is well reproduced by the model in the post-shock gas material, around $t\sim1000$~yr, when the gas has cooled down to $\sim 80\K$. The morphology of DCN together with the shock model suggests that gas-phase chemistry is the dominant mechanism producing the widespread DCN emission, which dominates in the head of the bow-shock.

\subsubsection{Nitrogen}

We do not find any evidence of  $^{15}$N isotopic fractionation.  Towards L1157-mm,  we obtain $^{13}$C/$^{15}$N ratios of $5.8\pm 2.2$ and $4.8\pm 2.4$ for HCN and HNC, respectively. Both values agree within the uncertainties, and they are also consistent with the canonical elemental abundance ratio of 6.5  in the local ISM \citep{Milam2005,Marty2011}.  This translates into  elemental abundance ratios HC$^{14}$N/HC$^{15}$N of $380\pm 145$ and H$^{14}$NC/H$^{15}$NC $\approx 320\pm 160$ towards L1157-mm. Both values are consistent  with the solar value in the local ISM (440; \citealt{Marty2011}). Towards L1157-B1, the H$^{13}$CN/HC$^{15}$N is almost unchanged ($5.0\pm 2.5$) whereas the  HN$^{13}$C/H$^{15}$NC takes a lower value ($3.0\pm 1.5$), but still in rough agreement with the canonical value. 


At the passage of the shock,  the rare $^{15}$N  isotopologues display the same behaviour as   HCN and HNC, as can be seen in Table~\ref{LTEfitting}. The H$^{13}$CN and HC$ ^{15}$N abundances in L1157-B1 increase by a factor of 140  with respect to the  values measured towards L1157-mm. 
By comparison, HN$^{13}$C and  H$ ^{15}$NC increase by  a factor ($\sim 8$). Overall, these variations remain very modest. 
Therefore, our analysis brings {\em direct} observational evidence  that the $^{14}$N/$^{15}$N elemental abundance ratio is not affected by the passage of the shock. 

Recently, \citet{Benedettini2021} have investigated the $^{14}$N/$^{15}$N fractionation of HCN in the two shocked clumps, B1 and B0, of the L1157 outflow using the NOEMA interferometer at $\sim 3\arcsec$ resolution, and obtained $^{14}$N/$^{15}$N$=340\pm70$, in good agreement with our determination (see Table~\ref{ratios}). Based on the chemical shock modelling, \citet{Benedettini2021} conclude that the rich gas chemistry activated by the passage of the shock does not affect the $^{14}$N/$^{15}$N ratio with respect to the local ISM value. This conclusion is observationally confirmed by our results. 

\begin{figure*}
\centering
\includegraphics[width=0.97\columnwidth]{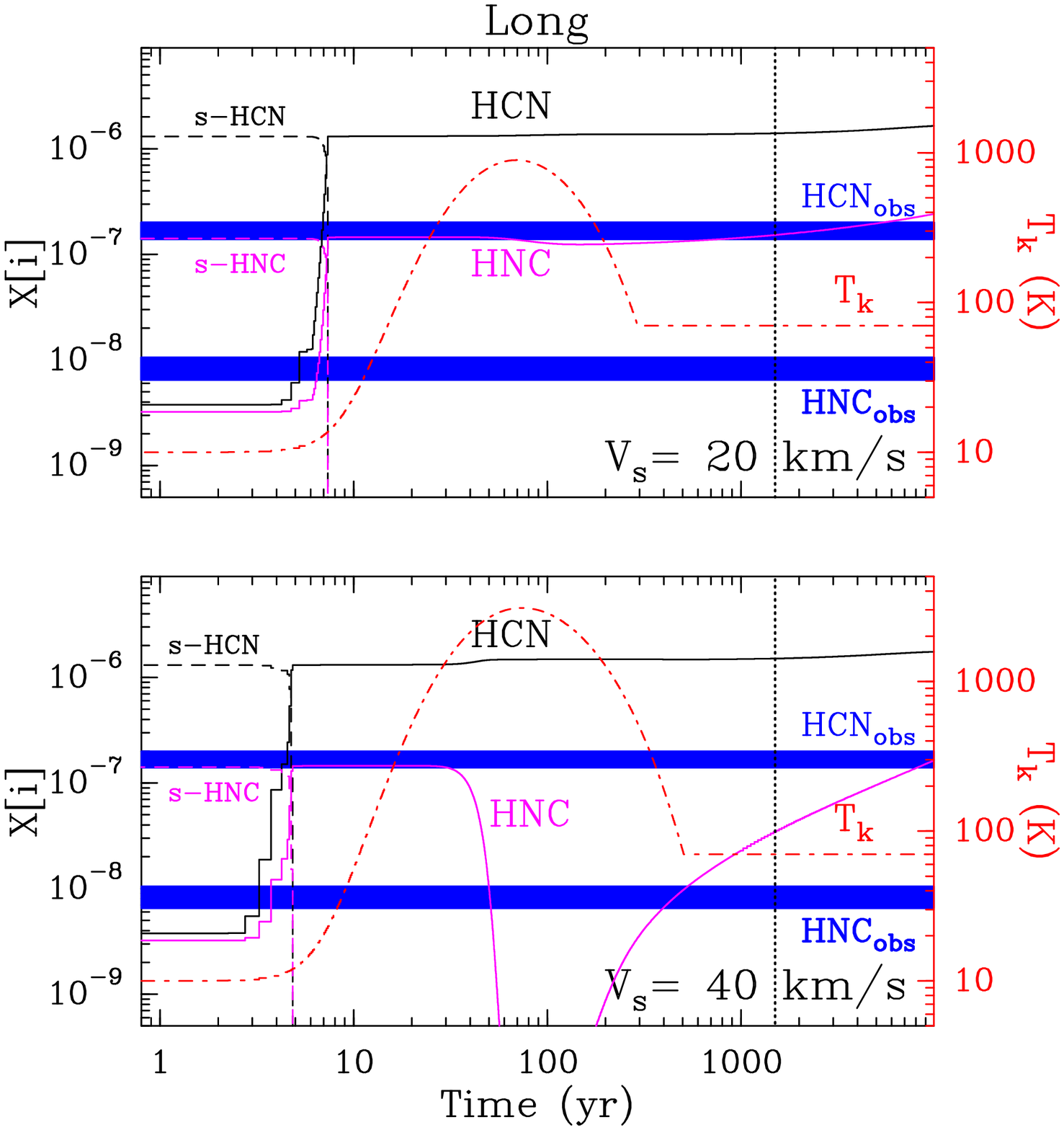}
\includegraphics[width=0.97\columnwidth]{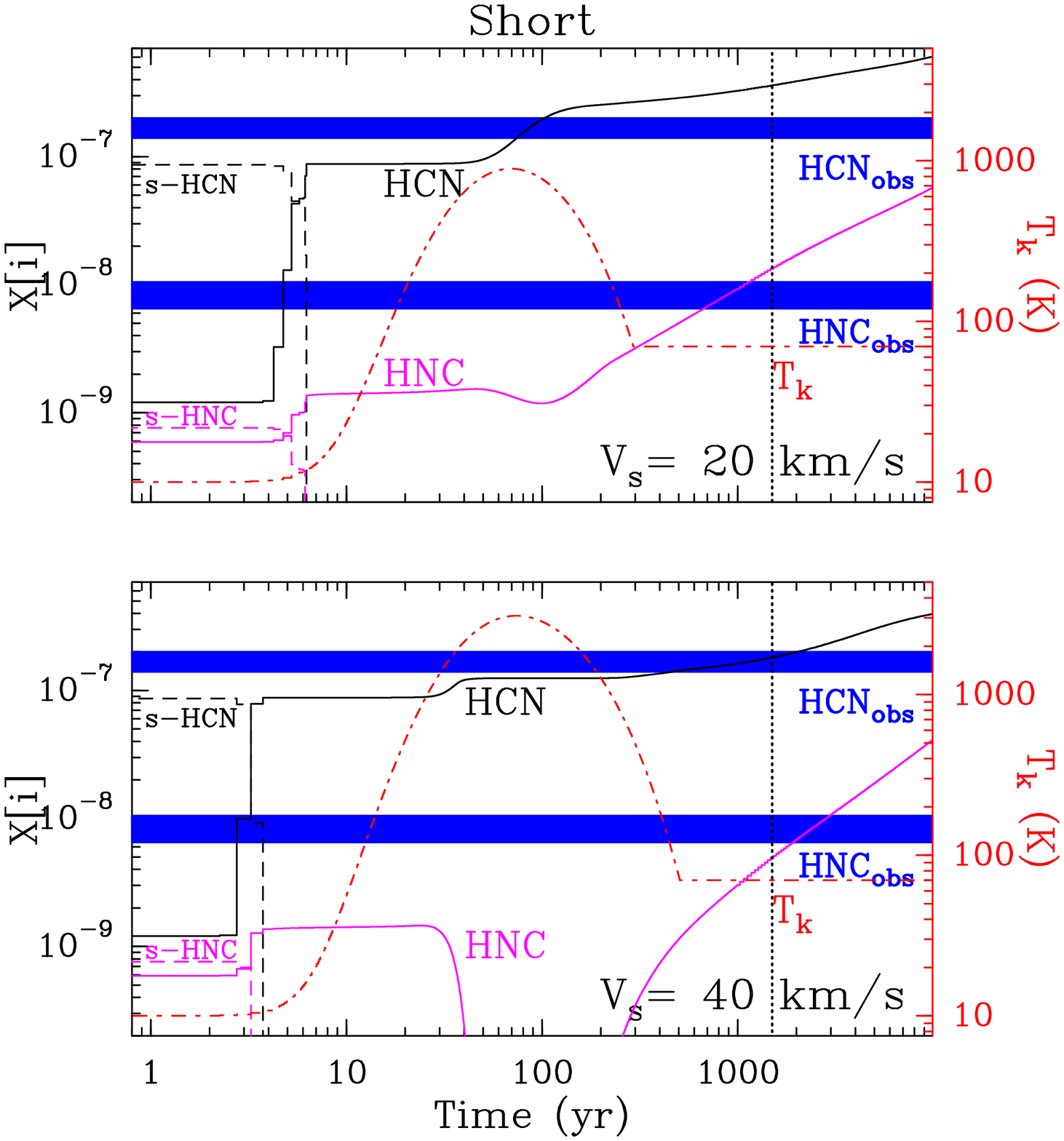}
\caption[]{ Time-dependent fractional abundances (with respect to the number of hydrogen nuclei) of HCN and HNC across a protostellar shock  propagating at $V=20\kms$ (top) and $V=40\kms$ (bottom), respectively. {\em (left)}"Long" pre-shock phase.  {\em (right)} "Short" pre-shock phase. The abundances of HCN (black) and HNC (magenta) in the gas phase and on the dust grains are indicated by the solid and the dashed curve, respectively. The temperature profile across the shock is drawn with the red dash-dot curve. The fractional abundances measured towards L1157-B1 and their uncertainties are indicated by the blue rectangles. The vertical dotted line at $t= 1500\yr$ indicates the estimated age of the shock by \citet{Podio2016}.}
\label{fig_model}
\end{figure*}

\subsection{Chemical modelling}

As discussed in the previous section, a simple qualitative analysis of the spectral line signature shows evidence for the shock impact on Nitrile gas composition. The HCN isotopologue abundances  have been deeply affected in the shock,  increasing by more than one order of magnitude, whereas the HNC isotopologues  seem to have barely noticed its presence. 

Our goal is  to  identify the chemical processes  responsible for the HCN and HNC gas phase abundances in L1157-B1,  before and during the passage of the shock. The different studies of H$_2$O, NH$_3$, CH$_3$OH, H$_2$CO, HC$_3$N, H$_2$S, NO, PN, PO by our group \citep{Viti2011, Lefloch2016, Holdship2017, Benedettini2013, Mendoza2018, Benedettini2013,Busquet2017, Codella2018,Benedettini2021} have led to a coherent picture of L1157-B1, in which the detected molecular emission is accounted for  by a C-type shock propagating into a  pre-shock medium of  density $n(\htwo)\simeq 5\times 10^4\cmmt$ and shock velocity $V_s$ in the range $20-40\kms$. We have adopted the same methodological approach as in our previous studies, and have used the public time dependent gas-grain chemical and parametrised  shock model UCLCHEM code \citep{Holdship2017} to  follow the evolution of the gas phase composition across a protostellar shock, adopting  a cosmic ray ionization rate of 
$\zeta=1.3\times10^{-16}\smu$, in agreement with \citet{Podio2014}. We refer the reader to the release paper of UCLCHEM for a detailed description of this public code \citep{Holdship2017}. Briefly, the code is run in two phases, where Phase I forms a dense core out of a diffuse, essentially atomic medium. An initial density of $100\cmmt$ is adopted here, for the diffuse medium. During this phase, gas-phase chemistry, freezing on to dust particles and subsequent surface processing, occurs. 

We have considered two cases, depending on the duration on the pre-shock phase. In the first case ("short'' duration), one considers the extreme case where the jet impacts the cloud as soon as the latter is formed; in the second case (long'' duration), one begins with a situation where 5 million years pass after the cloud is formed and before the shock arrives. The density at the end of Phase I is a free parameter, called from now on the pre-shock density. The second phase computes the time-dependent chemical evolution of the gas and dust during the passage of a magnetized (C-type) shock whose structure is at steady-state. 

Figure~\ref{fig_model}  presents the variations of the HCN and HNC molecular abundances {\em relative to the number of H nuclei} as a function of time through the shock, as predicted by our model for the shock velocities $V_s=20\kms$ (top) and $V_s=40\kms$ (bottom) and for the durations of the pre-shock phase, long (left) and short (right). The HCN (HNC) abundance in the gas and the solid phase  are drawn in solid and dotted black (magenta), respectively. In this figure, the shock is propagating to the left. Since the shock is propagating at steady state, the time variations of the different physical or chemical parameters  (temperature, density, abundances) can simply be converted into space variations, via the relation $z$= $V_s\times t$, hence giving access to the shock structure itself.  The observational determinations of the gas phase abundances of HCN and HNC and their uncertainties are indicated by the dark blue rectangles.

\subsubsection{Pre-shock phase}
Our modelling of the pre-shock phase  is consistent with the physical and chemical conditions in 
the envelope of L1157-mm, i.e., cold gas at $10\K$ of moderate density ($n(\htwo)= 10^5\cmmt$), for which we determined an abundance of $1.2\times 10^{-9}$ and $1.1\times 10^{-9}$ for HCN and HNC, respectively, relative to H. 
The key observational constraint here is that the abundances of HCN and HNC in the gas phase are similar. As can be seen in Fig.~\ref{fig_model}, our chemical modelling succeeds in reproducing similar gas phase abundances for HCN and HNC in the pre-shock phase. The HCN/HNC ratio barely varies between the long duration ($\sim 1$) and the short duration scenario ($\sim 2$).   Both scenarios also predict correct gas phase abundances relative to H, of the order of $10^{-9}$ (Table~\ref{LTEfitting}). The short duration scenario predicts gas phase abundances in slightly better agreement with our observational determinations, within the observational uncertainties. The "long" duration scenario predicts abundances higher by a factor of 3. The gas phase abundances of the rare isotopologues are also in good agreement with our observational determinations and with the absence 
of $^{13}$C  and $^{15}$N fractionation, as discussed above in Sect.~5.3. 

In the gas phase, HNC and HCN are  efficiently produced by the dissociative recombination (DR) of HCNH$^+$  with a branching ratio $\approx 1$ \citep{Herbst2000,Mendes2012}: 
$$ \rm HCNH^+ + e^- \rightarrow HNC + H, HNC+H .$$ Several other routes involving neutral-neutral reactions can substantially contribute to the formation of HCN, especially at higher densities (and hence at later times). The dominant route of "destruction" for both species is freeze-out on the dust grains.

We show in Fig.~\ref{fig_model} (in magenta dashed) the abundances of HCN and HNC on the dust grains (respectively s-HCN and s-HNC). The main difference between the short and the long duration Phase 1 models lies in fact in the solid HCN/HNC ratio, whereby the latter model predicts a ratio 
of $\sim$ 10 while the former  a ratio of more than 100. This is probably due to the very efficient formation route for HNC in the gas phase, which 
then readily freezes out. On the surface of grains, HCN can also continue to form via the efficient hydrogenation of CN, while HNC is not produced nor efficiently destroyed on the grains.  This explains the larger ratio of solid HCN/HNC compared to the gas phase HCN/HNC. Therefore, the duration of the pre-shock phase has a marked impact on the HCN and HNC abundance in the solid phase.

\subsubsection{Shock impact}

In the models with $V_s$= $20\kms$ (top right and left panels in Fig.~\ref{fig_model}),  the shock impact on the HCN and HNC abundances is essentially the same. In the magnetic precursor region, where ions and neutrals are kinematically decoupled,  all the material stored at the surface of the dust grains is sputtered and released into the gas phase. This leads to a quick increase of both the HCN and HNC abundances in the gas phase in a few years. Therefore, in the early years, the gas phase abundance of HCN and HNC is mainly determined by the abundance of both species on the dust grains and, at later times, by the different pre-shock gas composition between the long and short Phase I models, as well as by temperature-dependent  gas-phase chemistry. The main differences between the long-duration and the short-duration scenarii are the late time behaviours of  HCN and HNC, where both increase, in the latter case, much more sharply. The reactions that seem to contribute to these increases are the two neutral-neutral reactions of molecular \htwo\ with CN (for HCN) and of atomic N with CH$_2$ (for HNC). 
A full chemical analysis of this behaviour is beyond the scope of this work, especially as the age of L1157-B1 is estimated to be $\sim 1500\yr$ \citep{Podio2016} but the shorter Phase I duration certainly implies a richer gas phase at the beginning of Phase 2, and hence more abundant nitrogen and CN. This is because of a less effective freeze out during Phase I which would have not only depleted these two species but also hydrogenated them into NH$_3$ and HCN respectively.

The models with  higher shock velocities ($V_s$= $40\kms$) and hence higher maximum temperatures (see bottom panels in Fig.~\ref{fig_model}), display qualitatively similar evolution for HCN among the two models. Quantitatively, the "short" duration scenario predicts HCN and HNC abundances both in agreement with our observational determinations for  time $t\simeq 2000\yr$, which  is also consistent with the  estimated age of B1 ($\sim 1500\yr$; \citet{Podio2016}). On the contrary, the "long" duration scenario results into an overabundance of HCN (HNC)  in the shocked gas by a factor of 10 (5) with respect to our estimates, so that the HCN abundance remains essentially constant with time. In both the "long" and the "short" duration scenarii, the HNC abundance suddenly drops to values of a few $10^{-11}$ when the temperature rises above a threshold value of $\approx 2000\K$. This is probably due to the endothermic reaction C$^+$ + HNC which only becomes efficient at a temperature above $2000\K$. 

Therefore,  our modelling favors a scenario with a "short" duration pre-shock phase and a shock propagating at $V_s= 40\kms$ into pre-density gas $n(\htwo)= 5\times 10^4\cmmt$. It successfully accounts quantitatively for the dramatic increase of HCN abundance with respect to HNC on timescales consistent with the age of the shock.  The dramatic increase of HCN/HNC mainly appears as an effect of grain sputtering, which is already efficient in  low velocity shocks ($V_s= 20\kms$). This could explain why the HCN and HNC emission is so extended along the outflow, while other species, like e.g. NH$_2$CHO, require more stringent shock conditions to be formed \citep{Codella2017}. We note that these results are fully consistent with our previous studies of L1157-B1.


\section{Conclusions}\label{conclusions}

Using the IRAM 30m telescope in the framework of ASAI, we have carried out a comprehensive observational study of the HCN/HNC isotopologues in the protostellar shock region L1157-B1.  

Based on the observations of the envelope of the protostar L1157-mm, we find that the abundances of  the rare HCN/HNC isotopologues in the pre-shock gas  are similar to those found in cold dark and prestellar cores. We find a ratio HCN/HNC close to unity for all rare isotopologues  and we derive  an elemental abundance ratio  $^{14}$N/$^{15}$N close to its elemental value in the ISM. There is no evidence of elemental  fractionation, except for Deuterium. 

The impact of the shock is characterized by a strong increase of the HCN/HNC abundance ratio and of the HCN abundance relative to $\htwo$.  The $^{13}$C and $^{15}$N isotopologues display the same qualitative and quantitative behaviour as the main isopologues, i.e. the  elemental abundance ratios remain unchanged in  the passage of the  shock.  

A chemical modelling using UCLCHEM successfully accounts for the properties of HCN/HNC in the gas phase and their evolution in the shock. A very good agreement with the observations is obtained for a steady-state C-shock propagating  with velocity $V_s= 40\kms$ into a pre-shock medium of density $\rm n(H)=10^5\cmmt$ at a time ($\sim 2000\yr$) reasonably consistent with the age of the shock as estimated by \citet{Podio2016}. These results agree  with our previous UCLCHEM studies on the chemical composition of L1157-B1 \citep{Viti2011, Lefloch2016, Holdship2017, Benedettini2013, Mendoza2018, Benedettini2013,Busquet2017, Codella2018,Benedettini2021}. 

The HCN/HNC chemical evolution appears to be tighly connected to the history and the composition of the pre-shock phase. In the case of L1157-B1, it is mainly driven  by grain sputtering, a process  also efficient in lower velocity shocks ($20\kms$). We propose this could explain the wide spatial extent of the HCN and HNC emissions in the L1157-B1  outflow, while other species (e.g. NH$_2$CHO) require more stringent shock conditions to be formed. 

\section*{Data availability}
The ASAI data are publicly available under the IRAM Data Archive webpage at https://www.iram.fr/ILPA/LP007/. The other, complementary data underlying this article will be shared on reasonable request to the corresponding author.

\section*{Acknowledgements}
Based on observations carried out as part of  the Large Program ASAI (project number 012-12) with the IRAM 30m telescope.
IRAM is supported by INSU/CNRS (France), MPG (Germany) and IGN (Spain).  This work  was supported by : (i) a grant from 
LabeX Osug@2020 (Investissements d'avenir - ANR10LABX56), by the French National Research Agency in the framework of the Investissements d’Avenir program (ANR-15-IDEX-02), through the funding of the "Origin of Life" project of the Univ. Grenoble-Alpes; (ii) the ERC Horizon 2020 research and innovation programme  "The Dawn of Organic Chemistry" (DOC), grant agreement No 741002; (iii) the ERC Horizon 2020 ITN Project “Astro-Chemistry Origins” (ACO), grant agreement No 811312. SV acknowledges the European Research Council (ERC) Advanced Grant MOPPEX 833460
E.M. and JL acknowledge support from the Brazilian agency FAPESP (grant 2014/22095-6 and 2015/22254-0). G.B. is supported by the State Agency for Research (AEI) of the Spanish MCIU through the AYA2017-84390-C2-2-R grant. 


\end{document}